\documentclass[aps,rmp,twocolumn,eqsecnum]{revtex4}
\usepackage{amsmath,bm,epsfig}
\usepackage{natbib}
\newcommand{\B}[1]{{\bm{#1}}}
\newcommand{\C}[1]{{\mathcal{#1}}}
\usepackage[latin1]{inputenc}
\newcommand{\Sb}[1]{_{_{\text {#1}}}} 
\newcommand{\Sub}[1]{_{_{\text {#1}}}}
\def\pDR{{\% {\rm DR}}}
\newcommand{\beq}{\begin{equation}}
\newcommand{\eeq}{\end{equation}}
\newcommand{\bea}{\end{eqnarray}}

\def\Re{${\C R}\mkern-3.1mu e$}
\def\RE{{\C R}\mkern-3.1mu e}

\def\Re{\ensuremath{{\C R}\mkern-3.1mu e}}
\def\De{\ensuremath{{\C D}\mkern-3.1mu e}}
 \def\({\left(} \def\){\right)}
\let \= \equiv
\begin{document}
\title{{\it  Colloquium}: Theory of Drag Reduction by Polymers in
Wall Bounded Turbulence}
\author{Itamar Procaccia, Victor S. L'vov}
\affiliation{Dept. of Chemical Physics, The Weizmann Institute of
Science, Rehovot 76100, Israel}
\author{Roberto Benzi}
\affiliation{Dip. di Fisica and INFN, Universit\`a ``Tor
Vergata", Via della Ricerca Scientifica 1, I-00133 Roma, Italy}

\begin{abstract}
The flow of fluids in channels, pipes or ducts, as in any
other wall-bounded flow (like water along the hulls of ships
or air on airplanes) is hindered by a drag, which increases
many-folds when the fluid flow turns from laminar to
turbulent. A major technological problem is how to reduce
this drag in order to minimize the expense of transporting
fluids like oil in pipelines, or to move ships in the ocean.
It was discovered in the mid-twentieth century that minute
concentrations of polymers can reduce the drag in turbulent
flows by up to 80\%. While experimental knowledge had
accumulated over the years, the fundamental theory of drag
reduction by polymers remained elusive for a long time, with
arguments raging whether this is a ``skin" or a ``bulk"
effect. In this colloquium review we first summarize the
phenomenology of drag reduction by polymers, stressing both
its universal and non-universal aspects, and then proceed to
review a recent theory that provides a quantitative
explanation of all the known phenomenology. We treat both
flexible and rod-like polymers, explaining the existence of
universal properties like the Maximum Drag Reduction (MDR)
asymptote, as well as non-universal cross-over phenomena
that depend on the Reynolds number, on the nature of the
polymer and on its concentration. Finally we also discuss
other agents for drag reduction with a stress on the
important example of bubbles.
\end{abstract}
\maketitle

\tableofcontents
\section{Introduction}

The fact that minute concentrations of flexible polymers can reduce
the drag in turbulent flows in straight tubes was first discovered
by B.A. Toms and published in 1949  \cite{Toms}. This is an
important phenomenon, utilized in a number of technological
applications including the Trans-Alaska Pipeline System. By 1995
there were about 2500 papers on the subject, and by now there are
many more. Earlier reviews include those by
\cite{69Lumley,72Hoyt,73Landhal,75Virk,90McComb,90Gennes,00Sreeni},
and others. In this introductory chapter we present some essential
facts known about the phenomenon, to set up the theoretical
discussions that will form the bulk of this review.

It should be stated right away that the phenomenon of drag, which is
distinguished from viscous dissipation, should be discussed in the
context of wall-bounded flows. In homogeneous isotropic turbulence
there exists {\em dissipation} \cite{Lamb}, which at every point of
the flow equals $\nu_0 |\B \nabla\B U (\B r,t)|^2$ where $\nu_0$ is
the kinematic viscosity and $\B U (\B r,t)$ is the velocity field as
a function of position and time. The existence of a wall breaks
homogeneity, and together with the boundary condition $\B U=0$ on
the wall it sets a {\em momentum flux} from the bulk to the wall.
This momentum flux is responsible for the drag, since not all the
work done to push the fluid can translate into  momentum in the
stream-wise direction. In stationary condition all the momentum
produced by the pressure head must flow to the wall. Understanding
this \cite{04LPPT} is the first step in deciphering the riddle of
the phenomenon of drag reduction by additives, polymers or others,
since usually these agents  tend to {\em increase the viscosity}. It
appears therefore counter-intuitive that they would do any good,
unless one understands  that the main reason for drag-reduction when
polymers or other drag-reducing agents are added to a Newtonian
fluid is caused by reducing the momentum flux to the wall. The rest
of this review elaborates on this point and makes it quantitative.

Contrary to the preference of the engineering community, in which
arguments can rage about what is the   {\em mechanism} of drag
reduction, and whether these are hairpin vortices or other `things'
that do the trick, we will adhere to the parlance of the physics
community where a theory is tested by its quantitative
predictiveness. We thus base our considerations on the analysis of
model equations and their consequences. The criterion for validity
will be our ability to describe and understand in a quantitative
fashion all the observed phenomena of drag reduction, both universal
and non-universal, and the ability to predict the results of yet
unperformed experiments.  The reader will judge for himself the
extent of success of this approach.

\subsection{Universality of Newtonian mean velocity profile}
\begin{figure}
\centering \epsfig{width=.45\textwidth,file=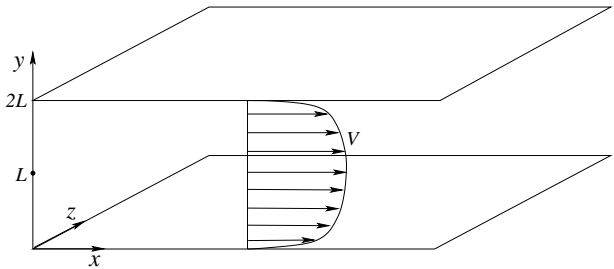}
\caption{The channel geometry}
\label{channel}
\end{figure}

\begin{figure}
\centering \epsfig{width=.49\textwidth,file=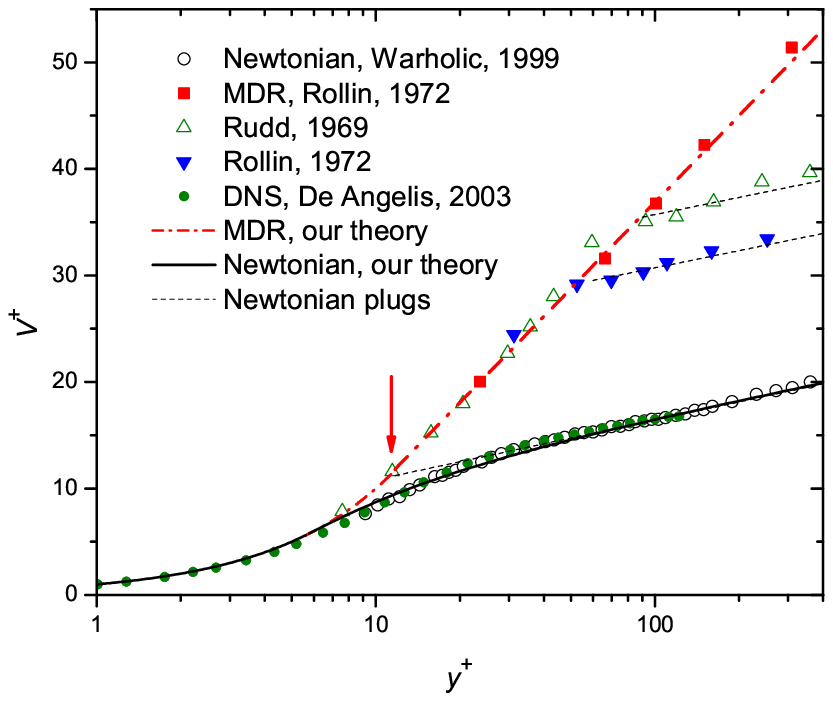} \caption{
Mean normalized  velocity profiles as a function of the normalized
distance from the wall. The data points (green circles)
 from numerical simulations of \textcite{03Angel}  and the
experimental points (open circles \cite{99War} refer to Newtonian
flows.  The solid line is a theoretical formula developed by
\textcite{04LPPT}. The red data points (squares) \cite{75Virk}
represent the Maximum Drag Reduction (MDR) asymptote. The dashed red
curve represents the log-law (\ref{finalexp}) which was derived from
first principles by \textcite{05BDLP}.  The blue filled triangles
\cite{72RS} and green open triangles \cite{69Rud} represent the
cross-over, for intermediate concentrations of the polymer,  from
the MDR asymptote to the Newtonian plug.} \label{profiles}
\end{figure}

For concreteness we will focus on channel flows, but most of our
observations will be equally applicable to other wall bounded flows.
The channel geometry is sketched in Fig. \ref{channel}.
The mean flow is in the $x$ direction, between two parallel plates
displaced by a distance $2L$. The distance from the lower wall is
denoted by $y$, and the span-wise direction by $z$. One takes the
length of the channel in the $x$ direction to be much larger than
the distance between the side walls in $z$, and the latter much
larger than $L$. In such a geometry the mean velocity $V \equiv
\langle \B U(\B r,t)\rangle$ is (to a high approximation)
independent of either $x$ or $z$, but only a function of the
distance from the wall, $V=V(y)$. When a Newtonian fluid flows at
large large Reynolds number (cf. Eq. (\ref{red}) in such a channel,
it exhibits in the near-wall region a {\em universal} mean velocity
profile. Here we use the word ``universal" in the sense that any
Newtonian fluid flowing in the vicinity of a smooth surface will
have the same mean velocity profile when plotted in the right
coordinates. The universality is best displayed in dimenisonless
coordinates, known also as ``wall units" \cite{Pope};  First, for incompressible fluids we can take
the density as unity, $\rho=1$. Then we define the
Reynolds number \Re, the normalized distance from the
wall $y^+$ and the normalized mean velocity $V^+(y^+)$ as follows:
\begin{equation}
\RE \equiv {L\sqrt{\mathstrut p' L}}/{\nu_0}\ , \  y^+
\equiv {y \RE }/{L} \ , \  V^+ \equiv
{V}/{\sqrt{\mathstrut p'L}} \ .\label{red}
\end{equation}
Here  $p'$ is the fixed pressure gradient $p'\equiv -\partial
p/\partial x$. The universal profile is shown in Fig. \ref{profiles}
with green circles [simulations of \textcite{03Angel}] open circles
(experiments) and a black continuous line  [theory by
\textcite{04LPPT}]. This profile has two distinct parts. For
$y^+\leq 6$ one observes the ``viscous sub-layer" where
\begin{equation}
V^+(y^+)=y^+ \ , \quad y^+\leq 6\label{vis}
\end{equation}
(and see Subsect. \ref{Karman} for a derivation), whereas for $y^+
\gtrsim 30$ one sees the celebrated
universal von K\'arm\'an ``log-law of the wall"
which is written in wall units as
\begin{equation}
V^+(y^+) =\kappa_{_{\rm K}}^{-1}\ln y^+ + B_{_{\rm K}}\,,  \quad{\rm
for}~ y^+ \gtrsim 30  \ . \label{Karman}
\end{equation}
 The law (\ref{Karman}) is universal, independent of the nature of
the Newtonian fluid; it had been a shortcoming of the theory of
wall-bounded turbulence that the von K\'arm\`an constant $\kappa_
{_{\rm K}}\approx 0.436 $ and the intercept $B_{_{\rm K}}\approx
6.13$ had been only known from experiments and simulations
\cite{79MY,97ZS}. Some recent progress on this was made by
\textcite{05LLPP}.

Having observed the universal profile for the mean velocity, it is
easy to see that any theory that seeks to understand drag reduction
by a change in viscosity \cite{90Gennes} is bound to fail, since the
universal profile is written in reduced coordinates, and a change in
the viscosity can result only in a re-parameterization of the
profile. Indeed, drag reduction must mean a change in the universal
profile, such that the velocity $V^+$ in reduced coordinates exceeds
the velocity $V^+$ predicted by Eq. (\ref{Karman}).
\subsection{Drag reduction phenomenology}
Here we detail some of the prominent features of drag reduction
\cite{75Virk}, all of which must be explained by a consistent
theory.  An immediate riddle that comes to mind is the following:
the polymers are molecular in scale. Turbulence is characterized by
the outer scale $L$ where energy is injected into the system, and by
the Kolmogorov  viscous scale $\eta$ below which viscous dissipation
dominates over inertial terms, and the velocity field becomes
essentially smooth. For all realistic flows the polymer size is much
smaller than this viscous scale (and cf. Subsec. \ref{FeneP} for
some actual numbers for a typical polymer). How is it then that the
polymers can interact at all with the turbulent degrees of freedom?
This riddle was solve by \textcite{69Lumley} who argued that it is
the polymer relaxation time $\tau$, the time that characterizes the
relaxation of a stretched polymer back to its coiled equilibrium
state, which is comparable to a typical eddy turn-over time in the
turbulent cascade. This matching of time scales allows an efficient interaction between the turbulent fluctuations and the polymer degrees of freedom. With the typical shear rate $S(y)$ one forms a
dimensionless ``Deborah number"
\begin{equation}
\De (y) = \tau S(y) \ . \label{defdeb}
\end{equation}
When \De~ exceeds the order of unity, the polymers begin to interact
with the turbulent flow by {\em stretching} and taking energy from
the turbulent fluctuations (cf. Subsec. \ref{FeneP} for a derivation
of this). We will see below that this mechanism of Lumley is
corroborated by all the available data. What we will have to explain
is how is it that as the polymers stretch, a process that must increase
the viscosity, nevertheless the drag reduces.

\subsubsection{The universal Maximum Drag Reduction asymptote}
One of the most significant experimental findings \cite{75Virk}
concerning turbulent drag reduction by polymers is that  in
wall-bounded turbulence (like channel and pipe flows) the velocity
profile (with polymers added to the Newtonian fluid) is bounded
between the von K\'arm\'an's log-law (\ref{Karman}) and another
log-law which describes the maximal possible velocity profile
(Maximum Drag Reduction, MDR),
\begin{equation}
V^+(y^+) = \kappa_{_{\rm V}}^{-1}\ln y^+ + B_{_{\rm V}}  \ ,
 \label{finalexp}
\end{equation}
where $ \kappa_{_{\rm V}}^{-1}\approx 11.7$ and $B_{_{\rm V}}\approx
-17$. This law, which was discovered experimentally by
\textcite{75Virk} (and hence the notation $\kappa_{_{\rm V}}$), is
also claimed to be universal, independent of the Newtonian fluid and
the nature of the polymer additive, including flexible and rod-like
polymers \cite{97VSW}. This log-law, like von K\'arm\'an's log-law,
contains two phenomenological parameters. In  \cite{04LPPT} it
was shown that in fact this law contains only one parameter, and
can be written in the form:
\begin{equation}
V^+(y^+) = \kappa_{_{\rm V}}^{-1}\ln\left(e\, \kappa_{_{\rm V}}
y^+\right)\, \quad{\rm
for}~ y^+ \gtrsim 12  \ , \label{final}
\end{equation}
where $e$ is the basis of the natural logarithm.
The deep reason for this simplification will be explained below. For
sufficiently high values of \Re, and sufficiently high concentration
of the polymer $c_p$, and length of polymer (number of monomers
$N_p$), the velocity profile in a channel is expected to follow the
law (\ref{final}). Needless to say, the first role of a theory of
drag reduction is to provide an explanation for the MDR law and for
its universality. We will explain below that the deep reason for the universality of the MDR is
that it is a marginal state between a turbulent and a laminar regime of wall bounded-flows. In this marginal state turbulent fluctuations almost do not contribute to the momentum and energy balance, and the only role of turbulence is to extend the polymers in a proper way.  This explanation can be found in Sect. \ref{MDR}, including an a-priori calculation of the parameter
$\kappa_{_{\rm V}}$. For finite \Re, finite concentration $c_p$, and
finite number of monomers $N_p$, one expects cross-overs that are
non-universal;  in particular such cross-overs depend on the nature
of the polymer, whether it is flexible or rod-like.
\subsubsection{\label{crossflex}
Cross-overs with flexible polymers} When the drag reducing agent is
a flexible polymer, but the concentration $c_p$ of the polymer is
not sufficiently large, the mean velocity profiles exhibit a
cross-over back to a log-law which is parallel to the law
(\ref{Karman}), but with a larger mean velocity (i.e. with a larger
value of the intercept $B_{_{\rm K}}$), see Fig.~\ref{profiles}. The
region of this log-law is known as the ``Newtonian plug". The
position of the cross-overs are not universal in the sense that they
depend on the nature of the polymers and the flow conditions. The
scenario is that the mean velocity profile follows the MDR up to a
certain point after which it crosses back to the Newtonian plug. The
layer of $y^+$ values between the viscous layer and the Newtonian
plug is referred to as the ``elastic layer". A theory for these
cross-overs is provided below. cf. Sect. \ref{Non}.

\begin{figure}
 \centering\epsfig{width=.50\textwidth,file=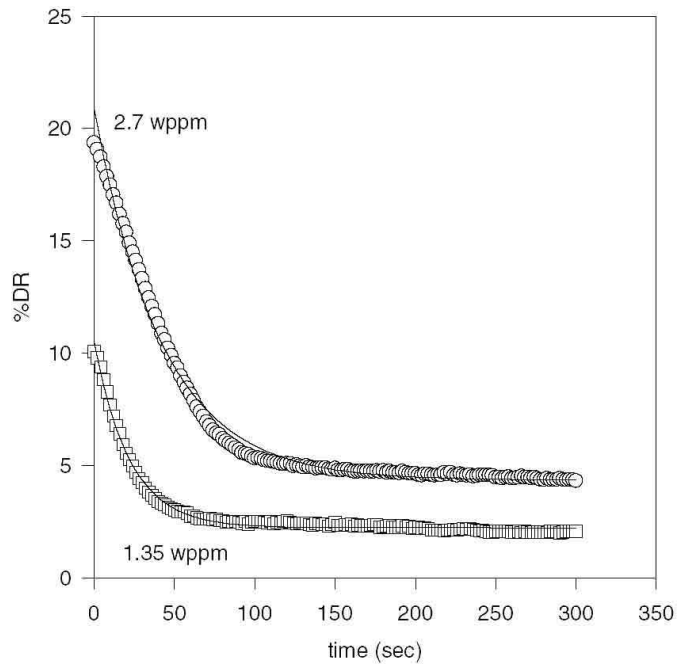}
\caption{$\%DR$ in a counter-rotating disks experiment with
$\lambda$-DNA as the drag reducing polymer. Note that the $\%DR$ is
proportional to $c_p$. When the length $N_ p$ reduces by a factor of
2 and, simultaneously, $c_ p$ increases by factor of $2$, the $\%DR$
reduces by a factor of 4. \label{Chan}}
\end{figure}

Another interesting experimental piece of information about
cross-overs was provided by \textcite{02CLLC}. Here turbulence was
produced in a counter-rotating disks apparatus, with $\lambda$-DNA
molecules used to reduce the drag. The Reynolds number was
relatively high (the results below pertain to Re~$\approx 1.2\times
10^6$) and the initial concentrations $c_p$ of DNA were relatively
low [results employed below pertain to $c_p=2.70$ and $c_p=1.35$
weight parts per million (wppm)]. During the experiment DNA
degrades; fortunately the degradation is very predictable: double
stranded molecules with 48 502 base pairs (bp) in size degrade to
double stranded molecules with 23 100 bp. Thus invariably the length
$N_ p $ reduces by a factor of approximately 2, and the
concentration $c\sb p$ increases by a factor of 2. The experiment
followed the drag reduction efficacy measured in terms of the
percentage drag reduction defined by
\begin{equation}\label{def-DR}
  \%DR = \frac{T_{_N} -T_{_V}}{T_{_N}}\times 100
\,,\end{equation}
 where $T_{_N}$ and $T_{_V}$ are the torques
needed to maintain the disk to rotate at a particular Reynolds
number  without and with polymers, respectively. The main
experimental results which are of interest to us are summarized in
Fig. \ref{Chan}.
We see from the experiment that both initially (with un-degraded
DNA) and finally (with degraded DNA) the $\%DR$ is proportional to
$c_ p$. Upon degrading, which amounts to decreasing the length $N_p$
by a factor of approximately 2 and, simultaneously increasing $c_ p$
by factor of $2$,
 $\%DR$ decreases by a factor 4. Explanations of these findings can
be found in Subsect.~\ref{DNAdeg}.

\subsubsection{Cross-overs with rod-like polymers}
\label{crossrod}
\begin{figure}
\hskip -1 cm
\centering \epsfig{width=.50\textwidth,file=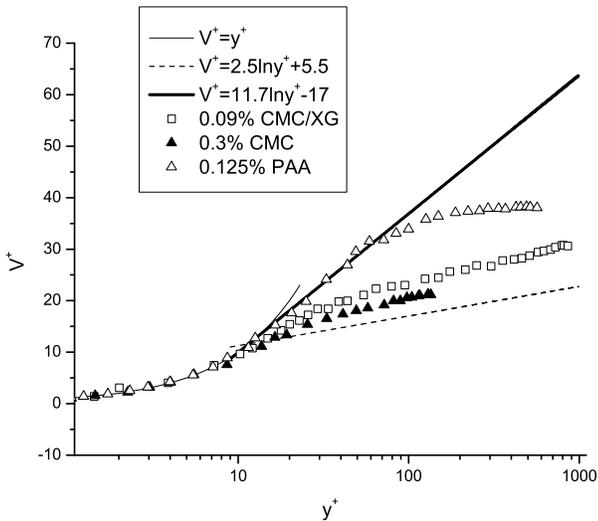}
\caption{Typical velocity profiles taken from \cite{Escudier}. In
dashed line we noted the von K\'arm\'an law (\ref{Karman}), while the
MDR (\ref{final}) is shown as the continuous black line. In all
cases the mean velocity follows the same viscous behavior for
$y^+<10$. After that the scenario is different for flexible and
rod-like polymers. The typical behavior for the former is presented
by the open triangles, which follow the MDR up to a cross-over
point that depends on the concentration of the polymer and on the
value of Re. The rod-like behavior is exemplified by the solid
triangles and the open squares; the mean velocity profiles appear
to interpolate smoothly between the two asymptotes as a function
of the concentration of the rod-like polymer.} \label{expt}
\end{figure}

 For the purposes of this review a flexible polymer is a polymer
that is coiled at equilibrium or in a flow of low Reynolds number,
and it undergoes a coil-stretch transition at some value of the
Reynolds number, (see below for details). A rod-like polymer is
stretched a-priori, having roughly the same linear extent at any
value of the Reynolds number. When the polymers are rod-like and the
concentration $c_p$ is not sufficiently high, the cross-over
scenario is different. The data in Fig. \ref{expt} include both
flexible and rod-like polymers \cite{Escudier}. It again indicates
that  for large values of \Re~the mean velocity profile with
flexible polymers [polyacrylamide (PAA)] follows the MDR until a
point of cross-over back to the ``Newtonian plug",  where it becomes
roughly parallel to von K\'arm\'an's log-law. Increasing the
concentration results in following the MDR further until a higher
cross-over point is attained back to the Newtonian plug.  On the
other hand, for rod-like polymers [sodium carboxymethylcellulose
(CMC) and sodium carboxymethylcellulose/xanthan gum blend (CMC/XG)]
the data shown in Fig. \ref{expt} indicate a different scenario.
Contrary to flexible polymers, here, as a function of the
concentration, one finds mean velocity profiles that interpolate
between the two asymptotes (\ref{Karman}) and (\ref{finalexp}),
reaching the MDR only for large concentrations. A similar difference
in the behavior of flexible and rod-like polymers when plotting the
drag as a function of Reynolds number was reported by
\textcite{97VSW}.
Clearly, an explanation of these differences between the way the MDR
is attained must be a part of the theory of drag reduction, and cf.
Sect. \ref{rodcross}. We reiterate that these cross-overs pertain to the situation in which  \Re~ is
large, but $c_p$ is too small.

\begin{figure}
\includegraphics[width=3.5in]{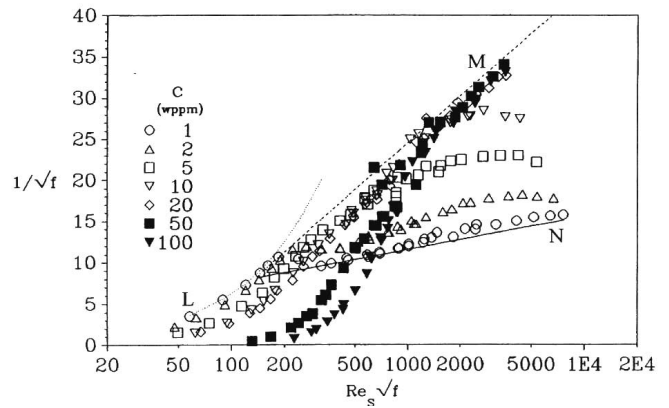}\\
  \caption{The drag in Prandtl-K\'arm\'an coordinates for a low concentration
 solution of NaCl in water mixed with the rod-like polymer PAMH B1120 in a pipe
  flow, see \cite{ASME} for details. The symbols represent the
  concentrations in wppm (weight parts per million) as given in the
  figure.  } \label{Virk}
\end{figure}

Another major difference between the two classes of polymers is
found when \Re~is too small, since then rod-like polymers can cause
drag {\em enhancement}, whereas flexible polymer never cause drag
enhancement. The latter are either neutral or cause drag reduction.
The best way to see the phenomenon of drag enhancement at low
\Re~with rod-like polymers is to consider the Fanning drag
coefficient defined as
\begin{equation}
f \equiv \tau_*/\Big(\frac{1}{2}\rho \tilde V^2\Big) \ ,
\end{equation}
where $\tau_*$ is the shear stress at the wall, determined by the value $S(y=0)$ of the shear
at $y=0$:
\begin{equation}
\tau_* \equiv \rho \nu_0 S(y=0) \ ,
\end{equation}
$\rho$ and $\tilde V$ are the fluid density and the mean fluid
throughput, respectively. Fig.~\ref{Virk} presents this quantity as
a function of \Re~in the traditional Prandtl-Karman coordinates
$1/\sqrt{f}$ vs. \Re$\sqrt{f}$, for which once again the Newtonian
high \Re~ log-law is universal, and for which also there exists and
MDR universal maximum \cite{75Virk,ASME}. The straight continuous
line denoted by `N' presents the Newtonian universal law. Data
points below this line are indicative of a drag enhancement, i.e.,
an increase in the dissipation due to the addition of the polymer.
Conversely, data points above the line correspond to drag reduction,
which is always bound by the MDR asymptote represented by the dashed
line denoted by `M'. This figure shows data for a rod-like polymer
(a polyelectrolyte in aqueous solution at very low salt
concentration) and shows how drag enhancement for low values of Re
crosses over to drag reduction at large values of Re \cite{ASME,
Bonn}. One of the results of  the theory presented below is that it
reproduces the phenomena shown in Fig. \ref{Virk} in a satisfactory
manner.
We reiterate that with {\em flexible} polymers the
situation is very different, and there is no drag enhancement at
any value of Re. The reason for this distinction will be made
clear below as well.

\section{Simple Theory of the von K\'arm\'an Law}
As an introduction to the derivation of the MDR asymptote for fluids
laden with polymers we remind the reader first how the von
K\'arm\'an log-law (\ref{Karman}) is derived. The derivation of the
MDR will follow closely similar ideas with the modifications due to
the polymers taken carefully into account.

Wall-bounded turbulence in Newtonian fluids is discussed \cite{79MY,
Pope} by considering the fluid velocity ${\B U}(\B r)$ as a sum of
its average (over time) and a fluctuating part:
\begin{equation}
\B U(\B r,t) =  V(y)\hat x + \B u(\B r,t)  \ . \label{split}
\end{equation}
The objects that enter the Newtonian theory are the mean shear $S(y)$, the
Reynolds stress $W(y)$ and the kinetic energy density per unit mass $K(y)$ :
\begin{equation}
S(y)\equiv d V(y)/d y \ , \!\!\!\quad
 W (y)\equiv - \langle u_x u_y\rangle
\ ,\! \!\!\quad K(y) = \langle |\B u|^2\rangle/2 \ . \label{SWK}
\end{equation}
Note that the Reynolds stress is nothing but the momentum in the
stream-wise direction $x$ transported by the fluctuations $u_y$ in
the direction of the wall; it is the momentum flux to the wall.
Using the Navier-Stokes equations one can calculate this momentum
flux $ P(y)$ which is generated by the pressure head $p'=-\partial
p/\partial x$;  at a distance $y$ from the wall this flux is
\cite{Pope}
\begin{equation}
P(y) = p'(L-y) \ . \label{P}
\end{equation}
We will be interested in positions $y\ll L$, so that we can
approximate the production of momentum by a constant $p'L$. This
momentum is then dissipated at a rate $\nu_0 S(y)$. This give rise
to the {\em exact} momentum balance equation
 \begin{equation}
\nu_0 S(y) +W(y) = p'L \  ,\quad y\ll L \ . \label{finalmom}
\end{equation}
The two terms on the LHS take their main roles at different values
of $y$. For $y$ close to the wall the viscous term dominates,
predicting that $S(y) \approx p'L/\nu_0$. This translates
immediately to Eq. (\ref{vis}), the mean velocity profile close to
the wall.

Away from the wall the Reynolds stress dominates, but all that can
be learned is that $W(y)\approx $constant, which is not enough to
predict the velocity profile. We need now to invoke a second
equation that describes the energy balance.

Directly from the Navier-Stokes equations one can compute the rate
of turbulent kinetic energy production by the mean shear, it is
$W(y) S(y)$ (see for example \textcite{Pope}), and the energy
dissipation $E_d$ at any point, $E_d=\nu_0 \langle |\nabla \B u(\B
r, t)|^2\rangle$. The energy dissipation is estimated differently
near the wall and in the bulk \cite{04LPPT}. Near the wall the
velocity is smooth and we can estimate the gradient by the distance
from the wall, and thus $E_d\approx \nu_0 a^2 K(y)/y^2$ where $a$ is
a dimensionless coefficient of the order of unity. Further away
from the wall the flow is turbulent, and we estimate the energy flux
by $K(y)/\tau(y)$ where $\tau(y)\approx  y/b\sqrt{K(y)}$ is the
typical eddy turn-over time at a distance $y$ from the wall, and $b$
is another dimensionless coefficient of the order of unity. Putting
things together yields the energy balance equation
\begin{equation}
 \Big[\nu_0\frac{a^2}{y^2}+ \frac{b\sqrt{K}}{y}\Big]K(y)= W(y)S(y) \  ,\label{Ebal}
\end{equation}
where the first term is dominant near the wall and the second in the
bulk. The interpretation of this equation is similar to the momentum
balance equation except that the latter is exact; the first term on
the LHS is the viscous dissipation, the second the energy flux to
the wall, and the RHS is the production.

To close the problem, the balance equations need to be supplemented
by a relation between $K(y)$ and $W(y)$. Dimensionally these objects
are the same, and therefore in the bulk [where one uses dimensional
analysis to derive the second term of (\ref{Ebal})], we expect that
these objects must be proportional to each other. Indeed,
experiments and simulations \cite{75Virk,Nieu} corroborate this
expectation and one writes  \cite{04LPPT,05LPT}
\begin{equation}
W(y)=c\Sub N^2 K(y) \ , \label{WK}
\end{equation}
with $c\Sub N$ being apparently $y$-independent outside the viscous
boundary layer. To derive the von K\'arm\'an log-law we now assert
that in the bulk the first term on the LHS of (\ref{Ebal}) is
negligible, we then use Eq. (\ref{WK}) together with the previous
conclusion that $W(y)=$const. to derive immediately $S(y)\propto
1/y$. Integrating yields the von K\'arm\'an log-law. Note that on
the face of it we have three phenomenological parameters i.e. $a$,
$b$ and $c\Sub N$. In fact in the calculation of the mean velocity
they appear in two combination which can be chosen as
$\kappa_{_{\rm K}}$ and $B_{_{\rm K}}$ in Eq. (\ref{Karman}).
Nevertheless it is not known how to compute these two parameters
a-priori.  Significantly, we will argue below that the slope of the
MDR can be computed a-priori, cf. Subsect. \ref{MDRth}.

\section{The Universality of the MDR}
\label{MDR} To study the implications of adding small concentrations
of polymers into wall-bounded turbulent flows we need a model that
describes reliably the modifications in the hydrodynamic equations
that are induced by the polymers. In the present section we are
aiming at the universal, model independent  properties of drag
reduction, and therefore any reasonable model of polymers
interacting with hydrodynamics which exhibits drag reduction should
also lead to the universal properties. We begin with the case of
flexible polymers.
\subsection{Model equations for flows laden with flexible-polymers}
\label{FeneP} 

To give a flavor of the type of polymers most commonly used in
experiments and technological applications of drag reduction we
present the properties of Polyethylene oxide
($N\times[\mbox{--CH}_2\mbox{--CH}_2\mbox{--O}]$, known as PEO). The
typical number $N$ of monomers ranges between $10^4$ and $10^5$. In
water solution in equilibrium this polymer is in a coiled state,
with an end-to-end distance $\rho_0$ of about $10^{-7}$ m. When
fully stretched the maximal end-to-end distance $\rho_m$ is about
$5\times 10^{-6}$ m. The typical mass concentration used is between
10 to 1000 wppm. The viscosity of the water solution increases by a
factor of 2 when PEO is added with concentration of 280 wppm.  Note
that even the maximal value value $\rho_m$ is much smaller than the
typical Kolmogorov viscous scale, allowing us to model the turbulent
velocity around such a polymer as a fluctuating homogeneous shear.

Although the polymer includes many monomers and therefore many
degrees of freedom, it was shown during years of research
\cite{Flory,87BCAH,94BE,Gen} that the most important degree of
freedom is the end-to-end distance, allowing the simplest model of
the polymer to be a dumbbell of two spheres of radius $\tilde a$ and negligible mass,
connected by a spring of equilibrium length $\rho_0$, characterized
by a spring constant $k$. This model allows an easy understanding of
the coil-stretch transition under a turbulent  shear flow. If such a
dumbbell is stretched to length $\rho_p$, the restoring force
$k(\rho_p-\rho_0)$ is balanced by the Stokes force
$6\pi \tilde a \nu_0 d\rho_p/dt$ (recall that the fluid density was taken as unity). Thus the relaxation time $\tau$ of the dumbbell is
$\tau=6\pi \tilde a \nu_0/k$.  On the other hand, in the presence of a
homogeneous shear $S$, the spring is stretched by a Stokes force
$6\pi \tilde a \nu_0 S\rho_p$.  Balancing the
stretching force with the restoring force (both proportional to $\rho_p$ for $\rho_p\gg \rho_0$), one finds that the coil-stretch transition is expected when
$S\tau>1$. In fact, a free polymer will rotate under a homogeneous
shear, making this argument a bit more involved. Indeed, one needs a
{\em fluctuating} shear $s$ in order to stretch the polymer, leading
finally to the condition
\begin{equation}
\De\approx \tau \sqrt{\langle s^2\rangle } >1 \ . \label{cost}
\end{equation}

To obtain a consistent hydrodynamic description of of polymer laden
flows one needs to consider a field of polymers instead of a single
chain. In a turbulent flow one can assume that the concentration of
the polymers is well mixed, and approximately homogeneous. Each chain still has
many degrees of freedom. However, as a
consequence of the fact that the most important degree of freedom
for a single chain is the end-to-end distance, the effects of the
ensemble of polymers enters the hydrodynamics in the form of a
``conformation tensor" $R_{ij}(r,t)$ which stems from the ensemble
average of the dyadic product of the end-to-end distance of the
polymer chains \cite{87BCAH,94BE}, $R_{\alpha\beta} \equiv \langle
\rho_\alpha \rho_\beta \rangle$. A successful model that had been
employed frequently in numerical simulations of turbulent channel
flows is the so-called FENE-P model \cite{87BCAH}, which takes into
account the finite extensibility of the polymers:
\begin{eqnarray}
\frac{\partial R_{\alpha\beta}}{\partial t}&+&( U_{\gamma}  \nabla_{\gamma})
R_{\alpha\beta}
=\frac{\partial U_\alpha}{\partial r_\gamma}R_{\gamma\beta}
+R_{\alpha\gamma}\frac{\partial U_\beta}{\partial r_\gamma}\nonumber\\
&&-\frac{1}{\tau}\left[ P(r,t) R_{\alpha\beta} -\rho_0^2
\delta_{\alpha\beta} \right]\ ,\label{EqR}\\
P(r,t)&&=(\rho_m^2-\rho_0^2)/(\rho_m^2 -R_{\gamma\gamma}) \ .
\end{eqnarray}
The finite extensibility is reflected by the Peterlin function
$P(r,t)$ which can be understood as a ensemble averaging correction
to the potential energy $k(\rho-\rho_0)^2/2$ of the individual
springs. For hydrodynamicists this equation should be evident: think
about magneto-hydrodynamics, and the equations for the magnetic
field $\B n$. Write down the equations for the diadic product
$n_\alpha n_\beta$; the inertial terms will be precisely the first
line of Eq. (\ref{EqR}). Of course, the dynamo effect would then
tend to increase the magnetic field, potentially without limit; the
role of the second line in the equation is to guarantee that the
finite polymer will not stretch without limit, and the Peterlin term
guarantees that when the polymer stretches close to the maximum
there will be rapid exponential decay back to equilibrium values of
the trace of $\B R$. Since in most applications $\rho_m\gg \rho_0$
the Peterlin function can also be written approximately as $P(
r,t)\approx (1/(1 -\gamma R_{\gamma\gamma})$ where
$\gamma=\rho_m^{-2}$. In its turn the conformation tensor appears in
the equations for fluid velocity $ U_{\alpha}( r,t)$ as an
additional stress tensor:
\begin{eqnarray}
&&\frac{\partial  U_{\alpha}}{\partial t}\!+\!( U_{\gamma}\nabla_{\gamma}) U_{\alpha}\!=\!-\nabla_{\alpha} p
+\nu_0 \nabla^2  U_{\alpha} + \nabla_{\gamma} {T_{\alpha\gamma}}\,,~ ~~\label{Equ}\\
&&{T_{\alpha\beta}}( r,t) = \frac{\nu_p}{\tau}\left[\frac{P(
r,t)}{\rho_0^2}  R_{\alpha\beta}(r,t) - \delta_{\alpha\beta} \right] \ . \label{Tfull}
\end{eqnarray}
Here $\nu_p$ is a viscosity parameter which is related to the
concentration of the polymer, i.e. $\nu_p/\nu_0\sim c_p$ where $c_p$
is the volume fraction of the polymer. We note however that the
tensor field can be rescaled to get rid of the parameter $\gamma$ in
the Peterlin function, $\tilde R_{\alpha\beta}=\gamma
R_{\alpha\beta}$ with the only consequence of rescaling the
parameter $\nu_p$ accordingly. Thus the actual value of the
concentration is open to calibration against the experimental data.
Also, in most numerical simulations, the term $P\rho_0^{-2} \B R$ is
much larger than the unity tensors in (\ref{EqR}) and (\ref{Tfull}).
Therefore, in the theoretical development below we shall use the
approximation
\begin{equation}
    T_{\alpha\beta} \sim\nu_p P R_{\alpha\beta}/\tau \rho_0^2 \ .
\label{T}
\end{equation}
We note that the conformation tensor always appears rescaled by
$\rho_0^2$. For notational simplicity we will absorb $\rho_0^2$ into
the definition of the conformation tensor, and keep the notation
$R_{\alpha\beta}$ for the renormalized, dimensionless tensor.

\begin{figure}
\centering\epsfig{width=.50\textwidth,file=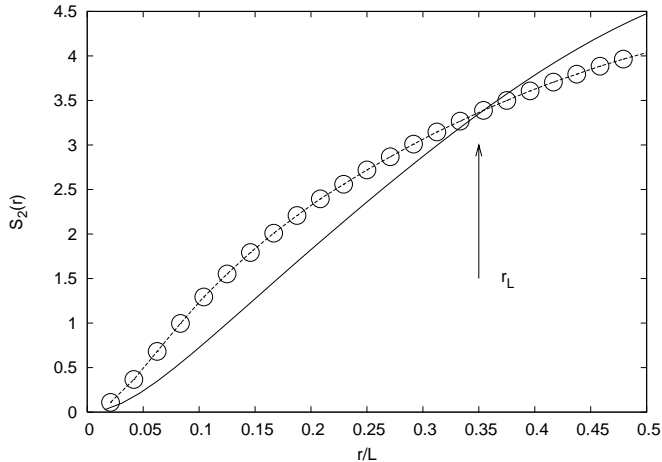}
\caption{Second order structure function $S_2(r)$ for homogeneous and isotropic turbulence.
Two cases are represented: with polymers (line with symbols) and without polymer (dashed line). The Lumley scale
is indicated by an arrow. The numerical simulations with polymers is performed using a $128^3$ resolution with
periodic boundary conditions. The energy content for scales below the Lumley scale is reduced indicating a significant energy transfer from the velocity field to the polymer elastic energy.}
\label{newfigure_benzi}
\end{figure}

Considering first homogeneous and isotropic turbulence, The FENE-P
model can be used to demonstrate the Lumley scale discussed above.
In homogeneous turbulence one measures  the moments of the velocity
difference across a length scale $r'$, and in particular the second
order structure function
\begin{equation}
S_2(r') \equiv\langle  \{[\B U(\B r+\B r',t)-\B U(\B r,t)]\cdot \frac{\B r'}{r'}\}^2\rangle \ ,
\end{equation}
where the average $\langle ... \rangle$ is performed over space and
time.  This quantity changes when the simulations are done with the
Navier-Stokes equations on the one hand and with the FENE-P
equations on the other, see \cite{DCBP}). The polymers decrease the
amount of energy at small scales. The reduction starts exactly at
the theoretical estimate of the Lumley scale, shown with an arrow in
Fig. \ref{newfigure_benzi}. A more detailed analysis of the energy transfer
\cite{DCBP} shows that energy flows from large to small scales and
at the Lumley scale a significant amount of energy is transferred to
the potential energy of the polymers by increasing $R_{\gamma
\gamma}$. This energy is eventually dissipated when the polymer relax their length
back to equilibrium. In contradistinction to the picture offered by  \textcite{90Gennes}, simulations indicate that the energy {\it never} goes
back from the polymers to the flow; the only thing that the polymers
can do is to increase the dissipation. This is a crucial statement
underlying again the challenge of developing a consistent theory of
drag reduction. In isotropic  and homogeneous conditions the effect
of polymer is just to lower the effective Reynolds number which, of
course, cannot explain drag reduction.

\begin{figure}
\includegraphics[width=9.5cm]{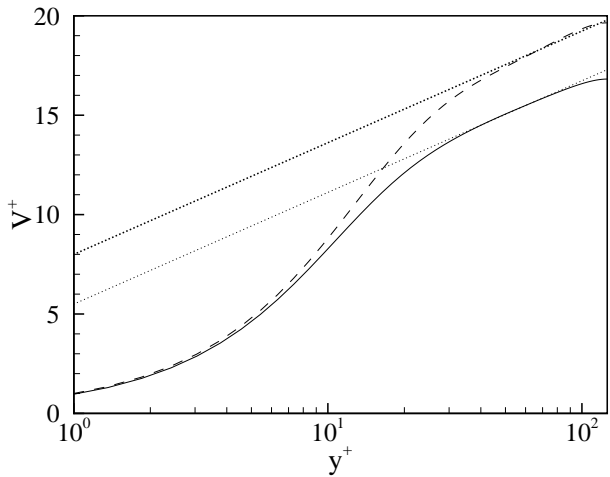}
\caption{ Mean velocity profiles for the Newtonian and for the
viscoelastic simulations with $\Re=125$ \cite{06BDLPT}.  Solid line:
Newtonian. Dashed line: Viscoelastic. The straight lines represent
a log-law with the classical von K\'arm\'an slope. Notice that in this
simulation the modest Reynolds number results in an elastic layer in the region
$y^+\le 20$.} \label{profile}
\end{figure}

The FENE-P equations were also simulated on the computer in a
channel or pipe geometry, reproducing the phenomenon of drag
reduction found in experiments \cite{98Dim,Nieu,06BDLPT,05Dim}. The
most basic characteristic of the phenomenon is the increase of fluid
throughput in the channel for the same pressure head, compared to
the Newtonian flow. This phenomenon is demonstrated in Fig.
\ref{profile}. As one can see the simulation is limited compared to
experiments; the Reynolds number is relatively low, and the MDR is
not attained. Nevertheless the phenomenon is there.
At any rate, once we are convinced that the FENE-P model exhibits
drag reduction, it must also reproduce the universal properties of
the phenomenon, and in particular the MDR. We will show that this is
indeed the case, but that also all the cross-over non-universal
phenomena can be understood using this model. If we were not
interested in the cross-over phenomena we could use directly the
large concentration limit of the FENE-P model, which is the limit
$P=1$ where the model identifies with the harmonic Dumbbell-model
known also as the Oldroid-B model.
\subsection{Derivation of the MDR}

In this section we review the theory that shows how the new log-law
(\ref{final}) comes about when the polymers are added to the flow.
In the next section we explain why this law is universal and
estimate the parameters from first principles.

As before, we proceed by taking the long time average of Eq.
(\ref{Equ}), and integrating the resulting equation along the $y$
coordinate. This produces an exact equation for the momentum
balance:
\begin{equation}
W + \nu_0 S + \frac{\nu\sb p}{\tau} \langle P R_{xy}\rangle (y) =
p'(L-y) \ . \label{mombalFene}
\end{equation}
The interpretation of this equation is as before, but we have a new
term which is  the rate at which momentum is transferred to the
polymers. Near the wall it is again permissible to neglect the term
$p'y$ on the RHS for $y\ll L$. One should not be surprised with the
form of Eq. (\ref{mombalFene}), the new term could only be the one
that is appearing there, since it must have an $x-y$ symmetry, and
it simply stems from the additional stress tensor appearing in Eq.
(\ref{Equ}). The derivation of the energy balance equation is more
involved, and had been described in great detail in
\cite{05LPPT,06BDLPT}. The final form of the equation is
\begin{equation}
a\nu_0\frac{K}{y^2}+ b\frac{K^{3/2}}{y}+c_4
\nu_p \langle R_{yy}\rangle  \frac{K(y)}{y^2} = WS \ ,
\label{balEvis}
\end{equation}
where $c_4$ is a dimensionless coefficient of the order of unity.
This equation is in its final form, ready to be analyzed further.
Equation (\ref{mombalFene}) needs to be specialized to the vicinity
of the MDR, which is only obtained when the concentration $c_p$ of
the polymers is sufficiently high,  when \Re~is sufficiently high,
but also when the Deborah number \De~ is sufficiently high. In
\cite{05LPPT,06BDLPT} it was shown that when these conditions are
met,
\begin{equation}
 \langle P R_{xy}\rangle = c_1\langle
R_{yy}\rangle S \tau \  , \label{xy}
\end{equation}
where $c_1$ is another dimensionless coefficient of the order of
unity. Using this result in Eq. (\ref{mombalFene}) we end up with
the momentum balance equation
\begin{equation}
W (y)+ \nu_0 S(y) + c_1 \nu\sb p \langle  R_{yy}\rangle (y) S(y) =
p' L \ ,\quad y\ll L \ . \label{mombalfinal}
\end{equation}
The substitution (\ref{xy}) is so important for what follows that we
dwell on it a bit further. When the conditions discussed above are
all met, the polymers tend to be stretched and well aligned with the
flow, such that the $xx$ component of the conformation tensor must
be much larger than the $xy$ component, with the $yy$ component
being the smallest, tending to zero with $\De\to \infty$. Since
\De~ is the only dimensionless number that can relate the various
components of $\langle R_{ij}\rangle$, we expect that $\langle
R_{xx}\rangle \sim \De \langle R_{xy}\rangle$ and $\langle
R_{xy}\rangle \sim \De \langle R_{yy}\rangle$. Indeed, a calculation
\cite{05LPPT} shows that in the limit $\De\to \infty$ the
conformation tensor attains the following universal form:
\begin{equation}
\langle \B R \rangle  \simeq \langle R_{yy}\rangle
\left(\begin{array}{ccc}
        2\De ^ 2 & ~\De~ &  ~~0~~\\
        \De & 1 & 0 \\
        0 & 0 & C
    \end{array} \right)\quad \mbox{for}\ \De\gg 1\ ,\label{Pi-a}
    \end{equation}
where $C$ is of the order of unity. We conclude that $R_{xx}$ is
much larger than any other component of the conformation tensor, but
it plays no direct role in the phenomenon of drag reduction. Rather,
$R_{yy}$ which is very much smaller, is the most important component
from our point of view.  We can thus rewrite the two balance
equations derived here in the suggestive form:
\begin{eqnarray}
\nu(y) S +W &= & p'L \  ,\label{finalmom}\\
\tilde a \nu(y) \frac{K}{y^2}+ b\frac{K^{3/2}}{y} &=&  WS \ , \label{effvisK}
\end{eqnarray}
where $\tilde a= a c_4/c_1$ and the ``effective viscosity" $\nu(y)$
is being identified as
\begin{equation}
\nu(y)\equiv \nu_0+ c_1 \nu_p \langle R_{yy} \rangle  \ . \label{effvis}
\end{equation}
Exactly like in the Newtonian theory one needs to add a
phenomenological relation between $W(y)$ and $K(y)$ which holds in
the elastic layer,
\begin{equation}
W(y)=c\Sub V^2 K(y) \ , \label{WKv}
\end{equation}
with $c\Sub V$ an unknown coefficient of the order of unity.

\begin{figure}
\includegraphics[width=0.5 \textwidth]{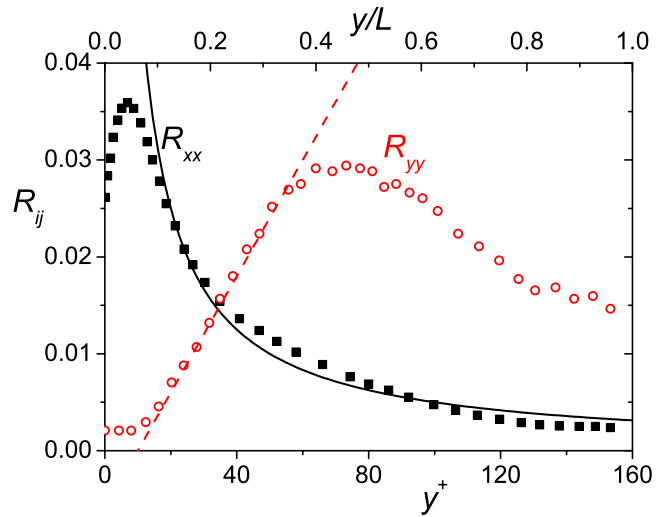}
\caption{\label{Rxx} Comparison of the DNS data Ref.~\cite{02SB}
for the mean profiles of $R_{xx}$ and $10\times R_{yy}$, the
components of the dimensionless conformation tensor, with analytical
predictions. In our notation $\Pi_0^{ij}\equiv \nu_{0p}R^2_{\rm max}
R_{ij}/(\tau_p R^2_{\rm eq})$. Black squares:
  DNS data  for the streamwise diagonal component $R_{xx}$, that
  according to our theory has to decrease as $1/y$ with the distance
  to the wall. Black solid line: the function $1/y^+$. Red empty
  circles: DNS data for the wall-normal component $10 R_{yy}$, for
  which we predicted a linear increase with $y^+$ in the log-law
  turbulent region. Red dashed line - linear dependence, $\propto
  y^+$. }
\end{figure}

At this point one asserts that the `dressed' viscous term dominates
the inertial term on the LHS of the balance equations
(\ref{finalmom}) and (\ref{effvisK}). From the first of these we
estimate $\nu(y)\sim p'L/S(y)$. Plugged into the second of these
equations this leads, together with Eq. (\ref{WK}) to
$S(y)=$Const.$/y$, where Const. is a combination of the unknown
coefficients appearing in these equations, and therefore is itself
an unknown coefficient of the order of unity. The important thing
is that the theory predicts a new log-law for $V(y)$, the slope of
which we will shown to be universal in the next Subsection. Before
we do so it is important to realize that if $S(y)=$Const.$/y$, our
analysis indicates that the effective viscosity $\nu(y)$ must grow
outside the viscous layer linearly in $y$, $\nu(y)\sim y$. If we
make the self-consistent assumption that $\nu_0$ is negligible in
the log-law region compared to $\nu_p \langle R_{yy}\rangle$, then
this prediction appears in contradiction with what is known about
the stretching of polymers in channel geometry, where it had been
measured that the extent of stretching {\em decreases} as a function
of the distance from the wall. The apparent contradiction evaporates
when we recall that the amount of stretching is dominated by
$R_{xx}$ which is indeed decreasing. To see this note that Eq.
(\ref{Pi-a}) predicts that $R_{xx}\approx 2 \De^2 R_{yy} \propto
1/y$. At the same time $R_{yy}$ increases linearly in $y$. Both
predictions agree with what is observed in simulations, see Fig.
\ref{Rxx}.

\begin{figure}
\includegraphics[width=0.45\textwidth]{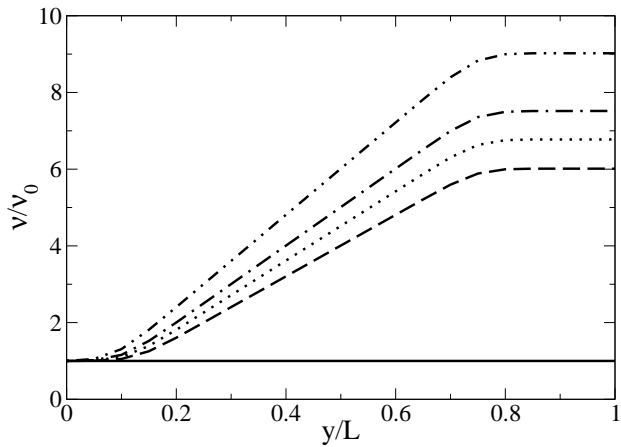}
\caption{The Newtonian viscosity profile and four examples of
close to linear viscosity profiles employed in the numerical
simulations. Solid line: run N, $-\, -$: run R, $\cdot \cdot \cdot
$: run S, $- \cdot -$: T, $- \cdot \cdot -$: run U. }
\label{visprof}
\end{figure}

\begin{figure}
\vskip 0.5 cm
\centering\includegraphics[width=0.45\textwidth]{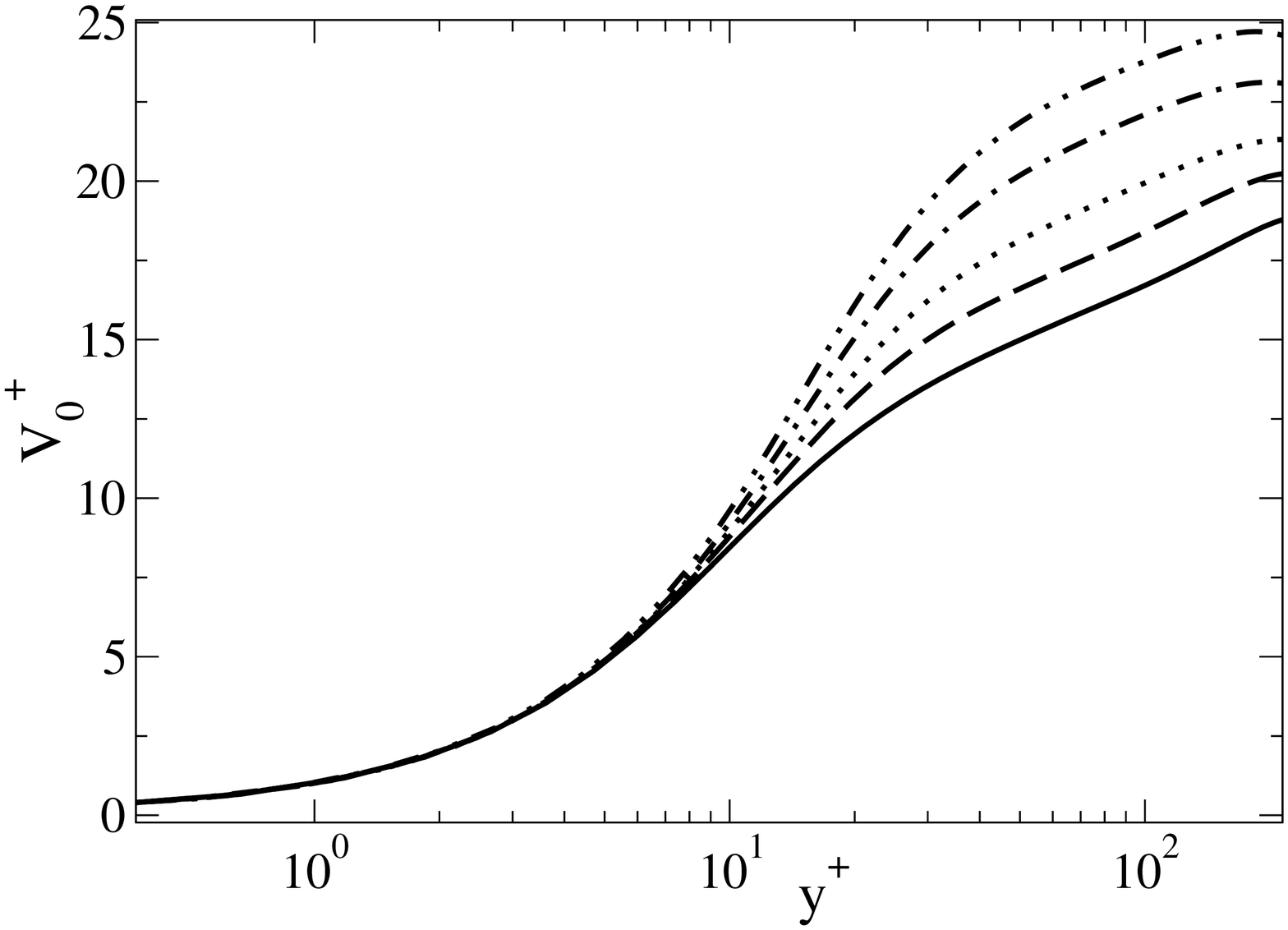}
\caption{The reduced mean velocity as a function of the reduced
distance from the wall. The line types correspond to those used in
Fig. \ref{visprof}.} \label{Vprofiles}
\end{figure}

The simplicity of the resulting theory, and the correlation between
a linear viscosity profile and the phenomenon of drag reduction
raises the natural question, is it enough to have a viscosity that
rises linearly as the function of the distance from the wall to
cause drag reduction? To answer this question one can simulate the
Navier-Stokes equations with proper viscosity profiles (discussed
below) and show that the results are in semi-quantitative agreement
with the corresponding full FENE-P DNS. Such simulations were done
\cite{04DCLPPT} in a domain $2\pi L\times2L\times 1.2\pi L$, with
periodic boundary conditions in the streamwise and spanwise
directions, and with no slip conditions on the walls that were
separated by 2$L$ in the wall-normal direction. An imposed mass flux
and the same Newtonian initial conditions were used. The Reynolds
number Re (computed with the centerline velocity) was $6000$ in all
the runs. The $y$ dependence of the scalar effective viscosity was
close to being piece-wise linear along the channel height, namely
$\nu=\nu_0$ for $y\le y_1$, a linear portion with a prescribed slope
for $y_1<y\le y_2$, and again a constant value for $y_2<y<L$. For
numerical stability this profile was smoothed out as shown in Fig.
\ref{visprof}.  Included in the figure is the flat viscosity profile
of the standard Newtonian flow.
In Fig. \ref{Vprofiles} we show the resulting profiles of
$V^+_0(y)$ vs. $y^+$. The line types are chosen to correspond to
those used in Fig. \ref{visprof}. The {\em decrease} of the drag
with the {\em increase} of the slope of the viscosity profiles is
obvious. Since the slopes of the viscosity profiles are smaller
than needed to achieve the MDR asymptote for the corresponding
\Re, the drag reduction occurs only in the near-wall region
and the Newtonian plugs are clearly visible.

The conclusion of these simulations appears to be that one can
increase the slope of the linear viscosity profile and gain further
drag reduction. The natural question that comes to mind is whether
this can be done without limit such as to reduce the drag to zero.
Of course this would not be possible, and here lies the clue for
understanding the universality of the MDR.

\subsection{The universality of the MDR: theory}
\label{MDRth} The crucial new insight that will explain the
universality of the MDR and furnish the basis for its calculation is
that {\bf the MDR  is a marginal flow state of wall-bounded
turbulence}: attempting to increase $S(y)$ beyond the MDR results in
the collapse of the turbulent solutions in favor of a stable laminar
solution with $W=0$ \cite{05BDLP}. As such, the MDR is universal by
definition, and the only question is whether a polymer (or other
additive) can supply the particular effective viscosity $\nu(y)$
that drives Eqs. (\ref{finalmom}) and (\ref{effvisK}) to attain the
marginal solution that maximizes the velocity profile. We expect
that the same marginal state will exist in numerical solutions of
the Navier-Stokes equations furnished with a $y$-dependent viscosity
$\nu(y)$. There will be no turbulent solutions with velocity
profiles higher than the MDR.

To see this explicitly, we first rewrite the balance equations in
wall units. For constant viscosity [i.e. $\nu(y) \equiv \nu_0$],
Eqs.  (\ref{finalmom})-(\ref{WK}) form a closed set of equations for
$S^+ \equiv S \nu_0/(p'L)$ and $W^+ \equiv W/p'L$ in terms of two
dimensionless constant ${\delta^+} \equiv a\sqrt{K/W}$ (the
thickness of the viscous boundary layer) and $\kappa_{_{\rm K}}
\equiv b/c_{_{\rm N}}^3$ (the Von Karman constant). Newtonian
experiments and simulations agree well with a fit using $\delta^+
\sim 6$ and $\kappa_{_{\rm K}}  \sim 0.436$ (see the black
continuous line in Fig. \ref{profiles} which shows the mean velocity
profile using these very constants). Once the effective viscosity
$\nu(y)$ is no longer constant we expect $c_{_{\rm N}}$ to change
($c_{_{\rm N}}\to c_{_{\rm V}}$) and consequently the two
dimensionless constants will change as well. We will denote the new
constants as $\Delta$ and $\kappa_{_{\rm C}} $ respectively. Clearly
one must require that for $\nu(y)/\nu_0 \rightarrow 1$, $\Delta
\rightarrow \delta^+$ and $\kappa_{_{\rm C}}  \rightarrow
\kappa_{_{\rm K}} $. The balance equations are now written as
\cite{05BDLP}:
\begin{eqnarray}
&&\nu^+(y^+) S^+ (y^+) +W^+(y^+)  =1\ , \label{bal1}\\
&&\nu^+(y^+)\frac{{\Delta}^2}{{y^+}^2
}+\frac{\sqrt{W^+}}{\kappa_{_{\rm C}}y^+} =S^+\ ,\label{bal2}
\end{eqnarray}
where  $\nu^+(y^+)\equiv \nu(y^+)/\nu_0$. Substituting now $S^+$
from Eq. (\ref{bal1}) into Eq. (\ref{bal2}) leads to a quadratic
equation for $\sqrt{W^+}$. This equation has as a zero solution for
$W^+$ (laminar solution) as long as $\nu^+(y^+){\Delta}/y^+=1$.
Turbulent solutions are possible only when $\nu^+(y^+){\Delta}/y^+
<1$. Thus at the edge  of existence of turbulent solutions we find
$\nu^+\propto y^+$ for $y^+ \gg 1$. This is not surprising, since it
was observed above that the MDR solution is consistent with an
effective viscosity which is asymptotically linear in $y^+$. It is
therefore sufficient to seek the edge solution of the velocity
profile with respect to linear viscosity profiles, and we rewrite
Eqs. (\ref{bal1}) and (\ref{bal2}) with an effective viscosity that
depends linearly on $y^+$ outside the boundary layer of thickness
$\delta^+$:
\begin{eqnarray}
&&[1+\alpha(y^+-\delta^+)]S^+ +W^+ =1\ ,\label{pr1}\\
&&[1+\alpha(y^+-\delta^+)]\frac{{\Delta}^2(\alpha)}{{y^+}^2
}+\frac{\sqrt{W^+}}{\kappa_{_{\rm C}}  y^+} =S^+\  . \label{pr2}
\end{eqnarray}

\begin{figure}
\centering \epsfig{width=.50\textwidth,file=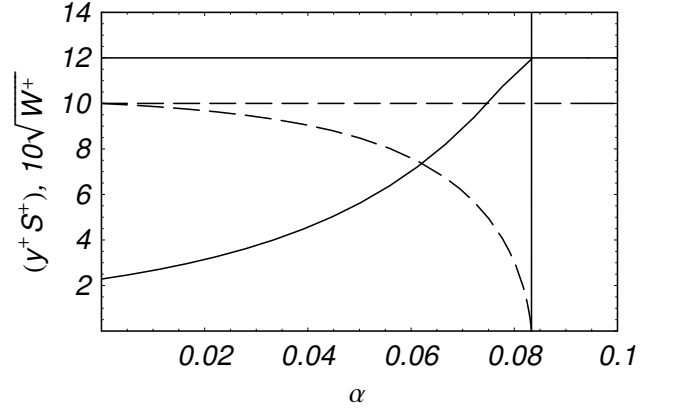}
\caption{The solution for 10$\sqrt{W^+}$ (dashed line) and $y^+S^+$
(solid line) in the asymptotic region $y^+\gg \delta^+$, as a
function of $\alpha$. The vertical solid line
$\alpha=1/2\delta^+=1/12$ which is the edge of turbulent solutions;
Since $\sqrt{W^+} $ changes sign here, to the right of this line
there are only laminar states. The horizontal solid line indicates
the highest attainable value of the slope of the MDR logarithmic
 law $1/\kappa_{_{\rm V}}=12.$         } \label{Fig1}
\end{figure}

We now endow $\Delta$ with an explicit dependence on the slope of
the effective viscosity  $\nu^+(y)$, $\Delta = \Delta(\alpha)$.
Since drag reduction must involve a decrease in $W$, we expect the
ratio $a^2 K/W$ to depend on $\alpha$, with the constraint that
$\Delta(\alpha)\to \delta^+$ when $\alpha\to 0$.  Although $\Delta$,
$\delta^+$ and $\alpha$ are all dimensionless quantities, physically
$\Delta$ and $\delta^+$ represent (viscous) length scales (for the
linear viscosity profile and for the Newtonian case respectively)
while $\alpha^{-1}$ is the scale associated to the slope of the
linear viscosity profile. It follows that $\alpha \delta^+$ is
dimensionless even in the original physical units. It is thus
natural to present $\Delta(\alpha)$  in terms of a dimensionless
scaling function $f(x)$,
\begin{equation}
\Delta(\alpha) =\delta^+ f(\alpha\delta^+) \ . \label{scaling}
\end{equation}
Obviously, $f(0)=1$. In the Appendix we show that the balance
equations (\ref{pr1}) and (\ref{pr2}) (with the prescribed form of
the effective viscosity profile) have an non-trivial symmetry  that
leaves them invariant under rescaling of the wall units.  This
symmetry dictates the function $\Delta(\alpha)$ in the form
\begin{equation}
\Delta(\alpha) =\frac{\delta^+}{1-\alpha\delta^+} \  . \label{Delta}
\end{equation}
Armed with this knowledge we can now find the maximal possible
velocity far away from the wall, $y^+\gg \delta^+$. There the
balance equations simplify to
\begin{eqnarray}
&&\alpha y^+S^+ +W^+ =1\ ,\label{pr12}\\
&&\alpha \Delta^2(\alpha)+
\sqrt{W^+}/\kappa_{_{\rm C}} =y^+S^+\  . \label{pr22}
\end{eqnarray}
These equations have the $y^+$-independent solution for $\sqrt{W^+}$ and
$y^+S^+$:
\begin{eqnarray}
\sqrt{W^+} &=&-\frac{\alpha}{2\kappa_{_{\rm C}}}
+\sqrt{\Big(\frac{\alpha}{2\kappa_{_{\rm C}}}\Big)^2
+1-\alpha^2\Delta^2(\alpha)} \ , \nonumber\\
y^+S^+&=&\alpha \Delta^2(\alpha)+\sqrt{W^+}/\kappa_{_{\rm C}} \ . \label{quadsol}
\end{eqnarray}
By using equation (\ref{quadsol}) , (see Fig. \ref{Fig1}), we obtain
that the edge solution ($W^+ \rightarrow 0$) corresponds to the
supremum of $y^+S^+$, which happens precisely when $\alpha=1/
\Delta(\alpha)$. Using Eq. (\ref{Delta}) we find the solution
$\alpha=\alpha_m =1/2\delta^+$.  Then $y^+S^+=\Delta(\alpha_m)$,
giving $\kappa^{-1}_{_{\rm V}}=2\delta^+$. Using the estimate
$\delta^+\approx 6$ we get the final prediction for the MDR. Using
Eq.  (\ref{final}) with $\kappa^{-1}_{_{\rm V}}=12$, we get
\begin{equation}
V^+(y^+) \approx 12\ln{y^+}  -17.8 \ . \label{predicta}
\end{equation}

This result is in close agreement with the empirical law
(\ref{finalexp}) proposed by Virk. The value of the intercept on the
RHS of Eq. (\ref{predicta}) follows from Eq. (\ref{final}) which is
based on matching the viscous solution to the MDR log-law in
\cite{04LPPT}. We now also have the deep justification for this
matching: the MDR is basically a laminar solution that can match
smoothly with the viscous sub-layer, with continuous derivative.
This is not possible for the von K\'arm\'an log-law which represents
fully turbulent solutions. Note that the numbers appearing in Virk's
law correspond to $\delta^+= 5.85$, which is well within the
error-bar on the value of this Newtonian parameter. Note that we can
easily predict where the asymptotic law turns into the viscous layer
upon the approach to the wall. We can consider an infinitesimal
$W^+$ and solve Eqs. (\ref{bal1}) and (\ref{bal2}) for $S^+$ and the
viscosity profile. The result, as before, is
$\nu^+(y)=\Delta(\alpha_m) y^+$. Since the effective viscosity
cannot fall bellow the Newtonian limit $\nu^+=1$ we see that the MDR
cannot go below $y^+=\Delta(\alpha_m)=2\delta^+$. We thus expect an
extension of the viscous layer by a factor of 2, in very good
agreement with the experimental data.

Note that the result $W^+=0$ should not be interpreted as $W=0$. The
difference between the two objects is the factor of $p'L$, $W= p'L
W^+$. Since the MDR is reached asymptotically as $\RE\to \infty$,
there is enough turbulence at this state to stretch the polymers to
supply the needed effective viscosity. Indeed our discussion is in
close correspondence with the experimental remark by Virk
\cite{75Virk} that close to the MDR asymptote the flow appears
laminar.

In summary,  the main message of this section is that the added
polymers endow the fluid  with an effective viscosity $\nu(y)$
instead of $\nu_0$.  There exists a profile of $\nu(y)$ that results
in a maximal possible velocity profile at the edge of existence of
turbulent solutions. That profile is the prediction for the MDR. In
particular we offer a prediction for simulations: direct numerical
simulations of the Navier-Stokes equations in wall bounded
geometries, endowed with a linear viscosity profile \cite{04DCLPPT},
will not be able to support turbulent solutions when the slope of
the viscosity profile exceeds the critical value that is in
correspondence with the slope of the MDR.

\section{The ``Additive Equivalence": the MDR of Rod-like Polymers}

In this section we address the experimental finding that {\em rigid
rod-like} polymers appear to exhibit the same MDR (\ref{final}) as
flexible polymers \cite{97VSW}. Since the bare equations of motion
of rod-like polymers differ quite significantly from those of
flexible polymers, one needs to examine the issue carefully to
understand this universality, which was termed by Virk ``Additive
Equivalence". We will see that the point is that in spite of the
different basic equations, when the conditions allow attainment of
the MDR, the balance equations for momentum and energy are identical
in form to those of the flexible polymers \cite{05BCLLP,06CLP}. The
differences between the two types of polymers arise when we consider
how the MDR is approached, and cross-overs back to the Newtonian
plugs, all issues that are taken up below.
\subsection{\label{rigid} Hydrodynamics with rod-like polymers}

The equation for the incompressible velocity field
$\B U (\B r,t)$ in the presence of rod-like polymers
has a form isomorphic to Eq. (\ref{Equ})
\begin{equation}\label{NS-a}
 \frac{\partial \B U}{\partial t}
  +\bm U  \cdot \bm\nabla \B U
= \nu_0\Delta\bm U -\bm\nabla p
    +\bm\nabla\cdot\bm\sigma \ ,
    \end{equation}
but with  another ${ \bm \sigma}
\Rightarrow \sigma_{ab}$ playing the role of an extra stress tensor caused by the
polymers.

The calculation of the  tensor $\bm \sigma$ for rigid rods is
offered in the literature \cite{88DE}, subject to the realistic
assumptions that the rod-like polymers are mass-less and having no
inertia. In other words, the rod-like polymers are assumed to be at
all times in local rotational equilibrium with the velocity field.
Thus the stress tensor does not have a contribution from the
rotational fluctuations against the fluid, but rather only from the
velocity variations along the rod-like object. Such variations lead
to ``skin friction", and this is the only extra dissipative effect
that is taken into account \cite{74Bre,75HL,76HL,03Man}. The result
of these considerations is the following expression for the
additional stress tensor:
\begin{equation}
\sigma_{ab} = 6\nu\sb p\,  n_a n_b \left( n_i n_j \C S_{ij}\right)
\,,\quad \mbox{rod-like polymers}\,,  \label{def-sigma}
\end{equation}
where $\nu\sb p$ is the polymeric contribution to the viscosity at
vanishingly small and time-independent shear;  $\nu\sb{p}$ increases
linearly with the polymer concentration, making it an appropriate
measure for the polymer's concentration.  The other quantities in
Eq.~(\ref{def-sigma}) are the velocity gradient tensor 
\begin{equation}
\C S_{ab}= \partial U_a/\partial x_b\,,
\end{equation}
 and $\bm{n}\equiv\bm{n}(\bm{r},t)$  is a unit
($\bm{n}\cdot\bm{n}\equiv1$) director field that describe the
polymer's orientation. Notice, that for flexible polymers the
equation (\ref{T}) for $T_{ab}$ is completely  different from
(\ref{def-sigma}). The difference between Eqs.~(\ref{def-sigma}) and
(\ref{T}) for the additional stress tensor in the cases of rod-like
and flexible polymers reflects their very different microscopic
dynamics. For the flexible polymers the main source of interaction
with the turbulent fluctuations is the stretching of the polymers by
the fluctuating shear $s$.  This is how energy is taken from the
turbulent field, introducing an additional channel of dissipation
without necessarily increasing the local gradient. In the rod-like
polymer case the dissipation is only taken as the skin friction
along the rod-like polymers. Having in mind  all these differences
it becomes even more astonishing that the macroscopic equations for
the mechanical momentum and kinetic energy balances  are isomorphic
for  the rod-like  and flexible polymers, as is demonstrated
below.
\subsection{The balance equations and the MDR}

Using Eq.~(\ref{def-sigma}) and Reynolds decomposition~(\ref{split}) we
compute
\begin{equation}\label{sigmarod}
\langle \sigma_{xy} \rangle = 6\nu\sb p \langle \C R_{xy}\C
R_{ij}\C S_{ij}\rangle = 6\, \nu_p[ S \langle \C R_{xy}^2\rangle + \langle  \C
R_{xy}\C R_{ij} s_{ij}\rangle ] \ .
\end{equation}
Now we make use of the expected solution for the conformation tensor
in the case of large mean shear. In such flows we expect a strong
alignment of the rod-like polymers along the streamwise direction
$x$. The director components $n_y$ and $n_z$ are then much smaller
than $n_x\approx 1$. For large shear we can expand $n_x$ according
to
\begin{equation}
\label{exp}
n_x=\sqrt{1-n_y^2-n_z^2}\approx 1-\frac12(n_y^2+n_z^2)\ .
\end{equation}
This expansion allows us to express  all products $ \C R_{ab} \C
R_{cd}=n_a n_b n_c n_d$ in terms that are linear in $\B{\C R}$, up
to third order terms in $n_y\sim n_z$. In particular
\begin{equation}\label{exm} \C R_{xx}^2\approx  1- 2(\C R_{yy}+\C
R_{zz})\,, \quad \C R_{xy}^2 \approx \C R_{yy} \ .
\end{equation}
With (\ref{exm}) the
first term in the RHS of  Eq. (\ref{sigmarod}) can be estimated as
\begin{equation}
\label{est3} 6\, \nu\sb p S \langle \C R_{xy}^2\rangle= \tilde c_1
\nu\sb p \langle R_{yy}\rangle S\,, \qquad \tilde c_1\simeq 6\ .
\end{equation}
The estimate of the second term on the RHS of Eq. (\ref{sigmarod})
needs some further calculations which can be found in
\cite{05BCLLP,06CLP}, with the result that it is of the same order
as the first one. Finally we can present therefore the momentum
balance equation in the form
\begin{equation}
\label{monabl} \nu_0 S + c_1\nu\sb p  \langle R_{yy}\rangle S + W  = p'L  \ .
\end{equation}
Another way of writing this result is in the form of an effective viscosity,
\begin{equation}
\label{eff}
\nu(y)  S + W  = p'L  \ ,
\end{equation}
where the effect of the rod-like polymers is included by the
effective viscosity $\nu(y)$:
\begin{equation}
\label{effnu}
\nu(y) \equiv \nu_0 + c_1\nu\sb p \langle R_{yy} \rangle \ .
\end{equation}
We see that despite the very different microscopic stress tensors,
the final momentum balance equation is the same for the flexible and
the rod-like polymers. Additional calculations which can be found in
\cite{05BCLLP} show that the energy balance equation also attains
the precisely the same form as Eq. (\ref{effvisK}). Evidently this
immediately translates, via the theory of the previous sections, to
the same MDR by the same mechanism, and therefore the ``additive
equivalence".

\section{Non-Universal Aspects of Drag Reduction: Flexible Polymers}
\label{Non} In this section we return to the cross-over phenomena
described in Subsect. \ref{crossflex} and Subsect. \ref{crossrod}
and provide the theory \cite{04BLPT} for their understanding. First
we refer to the experimental data in Fig. \ref{Chan}.

\subsection{The efficiency of drag reduction for flexible polymers}
When there exist cross-overs back from the MDR to the Newtonian plugs
\begin{figure}
\centering\includegraphics[width=0.45\textwidth]{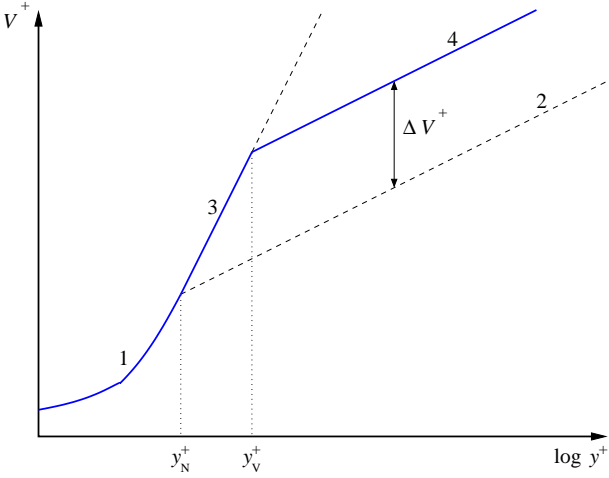}
\caption{Schematic mean velocity profiles. Region 1: $y^+ <
y^+\Sb N$, -- viscous sublayer.  Region 2: $ y^+ > y^+\Sb N $, --
logarithmic layer for the turbulent Newtonian flow. Region 3: $
y^+\Sb N  < y^+ < y^+ \Sb V =  (1+Q) \, y^+\Sb N$ -- MDR
asymptotic profile in the viscoelastic flow.  Region 4: $ y^+
> y^+ \Sb V $ -- Newtonian plug in the viscoelastic flow.
 \label{schemprof}}
\end{figure}
the mean velocity profile in the flexible polymer case  consists of
three regions \cite{75Virk}: a viscous sublayer, a logarithmic
elastic sublayer (region 3 in the Fig.~\ref{schemprof}) with the
slope greater then the Newtonian one, Eq. {\ref{finalexp}),
and a Newtonian plug (region 4). In the last region the velocity
follows a log-law with the Newtonian slope, but with some
 velocity increment $\Delta V^+$:
\begin{equation}\label{log-np}
  V^+(y^+) = \kappa\Sb K^{-1}\ln y^+ +B\Sb K+\Delta V^+ \
.\end{equation} Note that we have simplified the diagram for the
sake of this discussion: the three profiles Eqs. (\ref{vis}),
(\ref{Karman}), and (\ref{finalexp}) intercept at one point $y^+ =
y^+\Sb N\simeq\kappa\Sb V^{-1}\simeq11.7\approx 2\delta^+$. In
reality the Newtonian log-law does not connect sharply with the
viscous solution $V^+ = y^+$. but rather through a cross-over region
of the order of $\delta^+$.

The increment $\Delta V^+$ which determines the amount of drag
reduction is in turn determined by the cross-over from the MDR to
the Newtonian plug (see Fig.~\ref{schemprof}). We refer to this
cross-over point as $y^+\Sb V$. To measure the quality of drag
reduction one introduces \cite{04BLPT} a dimensionless drag
reduction parameter
\begin{equation}\label{Q}
 Q \equiv \frac{y^+\Sb V}{y^+\Sb N} -1\ .
\end{equation}
The velocity increment $\Delta V^+$ is related to this
parameter as follows
\begin{equation}\label{DeltaV}
  \Delta V^+ = \left(\kappa\Sb V^{-1}-\kappa\Sb K^{-1}\right)
    \ln\left(y^+\Sb V/y^+\Sb N\right)
  = \beta\ln(1+Q)
\ .\end{equation}
 Here $\beta\equiv\kappa\Sb V^{-1}-\kappa\Sb
K^{-1}\simeq9.4$. The Newtonian flow is then a limiting case of
the viscoelastic flow corresponding to $Q=0$.

The cross-over point $y^+\Sb V$ is non-universal, depending on \Re,
on the number of polymers per unit volume $c\sb p$, the chemical
nature of the polymer, etc. According to the theory of the last
Section, the total viscosity of the fluid $\nu_{\rm
tot}(y^+)=\nu_0+\nu_p(y^+)$ [where $\nu_p(y^+)$ is the polymeric
contribution to the viscosity which is proportional to $\langle
R_{yy}\rangle $] is linear in $y^+$ in the MDR region:
\begin{equation}\label{nu-tot}
  \nu_{\rm tot}(y^+) = \nu_0y^+/y^+\Sb N\,, \qquad
y^+\Sb N  < y^+ < y^+ \Sb V\ .
\end{equation}
When the concentration of polymers is small and \Re~ is large
enough, {\em the cross-over to the Newtonian plug at} $y^+\Sb V$
{\em occurs when the polymer stretching can no longer provide the
necessary increase of the total fluid viscosity}. In other words,
in that limit  the cross-over is due to the finite extensibility
of the polymer molecules. Obviously, the polymeric viscosity can
not be greater than $\nu_{p\max}$ which is the viscosity of the
fully stretched polymers. Thus the total viscosity is limited by
$\nu_0+\nu_{p\max}$. Equating $\nu_0+\nu_{p\max}$ and $\nu_{\rm
tot}(y^+\Sb V)$ gives us the cross-over position
\begin{equation}
  y^+\Sb V = y^+\Sb N(\nu_0+\nu_{p\max})/\nu_0
\ .\end{equation}
It follows from Eq. (\ref{Q}) that the drag reduction parameter is
determined very simply by
\begin{equation}\label{sigma}
  Q = \nu_{p\max}/\nu_0 \ ,
  \quad\text{$c\sb p$ small, \Re~ large} \ .
\end{equation}

At this point we need to relate the maximum polymeric viscosity
$\nu_{p\max}$ to the polymer properties. To this aim we estimate the
energy dissipation due to a single, fully stretched, polymer
molecule. In a reference frame co-moving with the polymer's center
of mass the fluid velocity can be estimated as $ u\simeq r \nabla u$
(the polymer's center of the mass moves with the fluid velocity due
to negligible inertia of the molecule). The friction force exerted
on the $i$-th monomer is estimated using Stokes law with $\delta
u_i$ being the velocity difference across a monomer,
\begin{equation}
F_i\simeq\rho_0\nu_0 \, a\,\delta u_i=\rho_0\nu_0 \,a\, r_i \nabla
u\, ,
\end{equation}
 where $a$ is an effective hydrodynamic radius of one monomer
(depending on the chemical composition), and $r_i$ is the distance
of the  $i$-th monomer from the center of the mass. In a fully
stretched state $r_i\simeq a\,i$ (the monomers are aligned along a
line). The energy dissipation rate (per unit volume) is equal to
the work performed by the external flow
\begin{eqnarray}
  -\frac{dE}{dt} &\simeq& c\sb p\sum_{i=1}^{N\sb p}
   F_i\delta u_i  \simeq
    \rho_0\nu_0a^3c\sb p N\sb p^3(\nabla u)^2 \nonumber\\
    &\equiv& \rho_0\nu_{p\max}(\nabla u)^2 \ .
\end{eqnarray}
We thus can estimate
$\nu_{p\max}$:
\begin{equation}
  \nu_{p\max} = \nu_0 a^3 c\sb p N\sb p ^3 \ . \label{result}
\end{equation}
Finally, the drag reduction parameter $Q$ is given by
\begin{equation}\label{sigma1}
  Q = a^3c\sb p N\sb p ^3 \qquad\text{$c\sb p$ small, \Re~ large}
\ .\end{equation} 
This is the central theoretical result of \textcite{04BLPT}, relating
the concentration $c\sb p$ and degree of polymerization $N\sb p $
to the increment in mean velocity $\Delta V^+$ via Eq.
(\ref{DeltaV}).

\subsection{Drag reduction when polymers are degraded}
\label{DNAdeg}

 The main experimental results which are of interest to us are
summarized in Fig. \ref{Chan}. Note that in this experiment the flow
geometry is rather complicated: with counter-rotating disks the {\em
linear} velocity depends on the radius, and the local  Reynolds
number is a function of the radius. The drag reduction occurs
however in a relatively  small near-wall region, where the flow can
be considered as a flow near the flat plate. Thus, one considers
\cite{04BLPT} an equivalent channel flow -- with the same \Re~
 and a  half width $L$ of the order of height/radius of the
cylinder. In this plane geometry the torques in (\ref{def-DR})
should be replaced by the pressure gradients $p'\Sb{N,V}$:
\begin{equation}\label{def-DR1}
  \pDR = \frac{p'\Sb N -p'\Sb V}{p'\Sb N}\times 100 \ .
\end{equation}
In order to relate $\pDR$ with the drag reduction parameter $Q$,
one re-writes Eq. (\ref{log-np}) in natural units
\begin{equation}
  V(y) =
    \sqrt{p'L}\left[
      \kappa_{_{\rm K}}^{-1}\ln \left(y\sqrt{p'L}/\nu_0\right)+B_{_{\rm K}}+\Delta V^+
    \right]
\ .\end{equation}
To find the degree of drag reduction one computes $p'_{_{\rm V}}$ and $p'_{_{\rm N}}$ keeping
the centerline velocity
$V(L)$ constant. Defining the centerline Reynolds number as Re, we rewrite
\begin{equation}\label{RE}
  {\rm Re} \equiv \frac{V(L) L }{\nu_0} =
    \RE\left[\kappa_{_{\rm K}}^{-1}\ln\RE +B_{_{\rm K}} +\Delta V^+\right]
\ .\end{equation}
 This equation implicitly determines the pressure gradient and
therefore the $\pDR$ as a function of $Q$ and Re. The set of
Eqs.~(\ref{DeltaV}) and (\ref{RE}) is readily solved numerically,
and the solution for three different values of Re~is shown in
Fig.~\ref{f:DR}. The middle curve corresponds to
Re$=1.2\times10^6$, which coincides with the experimental
conditions \cite{02CLLC}. One sees, however, that the dependence
of $\pDR$ on  Re is rather weak.
\begin{figure}
\centering\vskip.5cm\includegraphics[width=0.45\textwidth]{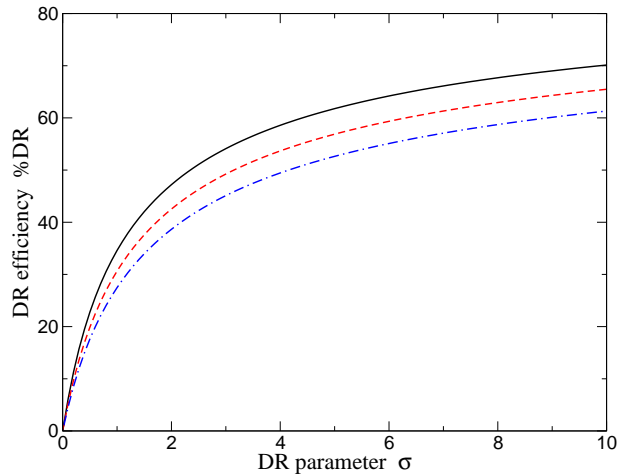}
\caption{Drag reduction efficiency $\pDR$ as function of the drag reducing
  parameter $Q$ for different centerline Reynolds numbers Re: $1.2\times10^5$,
  $1.2\times10^6$, and $1.2\times10^7$ (from top to bottom).\label{f:DR}}
\end{figure}
One important consequence of the solutions shown in Fig.~\ref{f:DR}
is that for small $Q$ (actually for $Q\le 0.5$ or $\pDR\le20$),
$\pDR$ is approximately a linear function of $Q$. The experiments
\cite{02CLLC} lie entirely within this linear regime, in which we
can linearize Eq.~(\ref{RE}) in $\Delta V^+$, solve once for
$p'_{_{\rm V}}$ and once for $p'_{_{\rm N}}$ (using $\Delta V^+=0$).
Computing Eq. (\ref{def-DR1}) we find an approximate solution for
the $\pDR$:
\begin{equation}\label{DR-lin}
  \pDR =
    \frac{2\beta Q}{\kappa\Sb K^{-1}\ln(\RE\Sb N) +B_{_{\rm K}}}\times100
\ .\end{equation}
Here $\RE \Sb N$ is the friction Reynolds number for the Newtonian flow,
i.e. the solution of Eq.~(\ref{RE}) for $\Delta V^+=0$.

It is interesting to note, that while the $\pDR$ depends on the
Reynolds number, the ratio of different $\pDR$'s does not [to $O(
Q)$]:
\begin{equation}\label{DRrel}
  \frac{\pDR^{(1)}}{\pDR^{(2)}} = \frac{Q^{(1)}}{Q^{(2)}} =
\frac{\nu_{p\max}^{(1)}}{\nu_{p\max}^{(2)}} \ .\end{equation} This
result, together with Eq. (\ref{result}), rationalizes completely
the experimental finding of \cite{02CLLC} summarized in Fig.
\ref{Chan}. During the DNA degradation, the concentration of
polymers increases by a factor of 2, while the number of monomers
$N\sb p $ decreases by the same factor. This means that $\pDR$
should decrease by a factor of 4, as is indeed the case.

The experimental results pertain to high \Re~and small $c\sb p$,
where we can assert that \emph{the cross-over results from
exhausting the stretching of the polymers such that the maximal
available viscosity is achieved}. In the linear regime that pertains
to this experiment the degradation has a maximal effect on the
quality of drag reduction $Q$, leading to the precise factor of 4 in
the results shown in Fig. \ref{Chan}. Larger values of the
concentration of DNA will exceed the linear regime as is predicted
by Fig. \ref{f:DR}; then the degradation is expected to have a
smaller influence on the drag reduction efficacy. It is worthwhile
to test the predictions of this theory also in the nonlinear regime.
\subsection{Other mechanism for cross-over}
Having any reasonable model of polymer-laden flows at our disposal
we can address other possibilities for the saturation of drag
reduction. We expect a cross-over from the MDR asymptote back to the
Newtonian plug when the basic assumptions on the relative importance
of the various terms in the balance equations lose their validity,
i.e. when (i) turbulent momentum flux $W$ becomes comparable with
the total momentum flux $p'L$, or when (ii) turbulent energy flux
$bK^{3/2}/y$ becomes of the same order as turbulent energy
production $WS$. In fact it was shown \cite{06BDLPT} using the
FENE-P model  that both these conditions give the same cross-over
point
\begin{equation}\label{co1}
  y\Sub V \simeq \frac{\tau\sqrt{p'L}}{\langle P\rangle} \
.\end{equation} Note that $\tilde\tau(y)\equiv\tau/\langle
P(y)\rangle$ is the effective non-linear polymer relaxation time.
Therefore condition (\ref{co1}) can be also rewritten as
\begin{equation}\label{co2}
  S(y\Sub V)\tilde\tau(y\Sub V) \simeq 1 \ .\end{equation} In
writing this equation we use the fact that the cross-over point
belongs also to the edge of the Newtonian plug where $S(y) \approx
\sqrt{p'L}/y$. The LHS  of this equation is simply the {\em local
Deborah number} (product of local mean shear and local effective
polymer relaxation time). Thus, {\em the cross-over to the Newtonian
plug ocurrs at the point, where the local Deborah number decreases
to $\sim1$}. We expect that this result is correct for any model of
elastic polymers, not only for the FENE-P model considered here.

To understand how the cross-over point $y\Sub V$ depends on the
polymer concentration and other parameters, one needs to estimate
mean value of the Peterlin function $\langle P\rangle$. Following
\textcite{06BDLPT} we estimate the value of $\langle P \rangle$ as
\begin{equation}
\langle P \rangle = \frac{1}{1-\gamma \langle R\rangle }\,,
\end{equation}
where $\langle R\rangle = \langle R_{xx}+R_{yy}+R_{zz}\rangle \sim
\langle R_{xx}+2R_{yy}\rangle $ and $\gamma \sim 1/\rho_m^2$ (for
simplicity we disregard $\rho_0$). We know from before that
\begin{equation}
  \langle R_{xx}\rangle \simeq (S\tilde\tau)^2 \langle R_{yy}\rangle
\,
\end{equation}
and therefore at the cross-over point (\ref{co2})
$$
  \langle R_{xx}\rangle \simeq \langle R_{yy}\rangle
\,,\quad
  \langle R\rangle \simeq \langle R_{yy}\rangle
\ .$$
The dependence of $\langle R_{yy}\rangle$ on $y$ in the MDR region follows from
Eqs.~(\ref{mombalfinal}), since
$$
  \langle R_{yy}\rangle \simeq
    \frac{y\sqrt{p'L}}{\nu\sb p}
\ .$$
Then at the cross-over point $y=y\Sub V$:
$$
 \langle P\rangle \simeq \frac1{1-\gamma y\Sub V\sqrt{p'L}\big/\nu\sb p}
\ .$$
Substituting this estimate into (\ref{co1}) gives the final result
\begin{equation}\label{co3}
 y\Sub V = \frac{C\tau\sqrt{p'L}}{1+\gamma p'L\tau\big/\nu\sb p} \
.\end{equation} Here $C$ is constant of the order of unity. Finally,
introducing dimensionless concentration of polymers
\begin{equation}
  \tilde c\sb p \equiv \frac{\nu\sb p}{\gamma\,\nu_0}
\,,\end{equation}
one can write denominator in Eq.~(\ref{co3}) as
$$
 1 +\frac{\gamma\,p'\,L\,\tau}{\nu\sb p} =
   1 +\frac{1}{\tilde c\sb p}\frac{p'\,L\,\tau}{\nu_0} =
   1 +\frac{\De}{\tilde c\sb p}
\,,$$
where

\begin{equation}
  \De \equiv \frac{p'\,L\,\tau}{\nu_0}
\end{equation}
is the (global) Deborah  number. Then for the dimensionless
cross-over point
$y\Sub V^+\equiv y\Sub V\sqrt{p'L}\big/\nu_0$ one obtains
\begin{equation}
  y\Sub V^+ = \frac{C \De}{1+De/\tilde c_p} \ . \label{crossfi}
  \end{equation}

This prediction can be put to direct test when $\tilde c_p$ is very
large, or equivalently in the Oldroyd B model where $P\equiv 1$
\cite{04BCHP}. Indeed, in numerical simulations when the Weissenberg
number was changed systematically, cf. \cite{01YKTM}, one observes
the cross-over to depend on \De~in a manner consistent with Eq.
(\ref{crossfi}). The other limit when $\tilde c_p$ is very small is
in agreement with the linear dependence on $\tilde c_p$ predicted in
Eq. (\ref{sigma1}).

We can thus reach conclusions about the saturation of drag
reduction in various limits of the experimental conditions, in
agreement with experiments and simulations.
\section{Cross-over Phenomena with Rod-like Polymers \label{rodcross}}

\subsection{The attainment of the MDR as a function of concentration}

At this point we return to the different way that the MDR is
approached by flexible and rod-like polymers when the concentration
of the polymer is increased. As discussed above, in the case of
flexible polymers the MDR is followed until the cross-over point
$y^+_v$ that was discussed already in great detail. The rod-like
polymers attain the MDR only asymptotically, and for intermediate
values of the concentration the mean velocity profile increase
gradually from the von K\'arm\'an log-law to the MDR log-law.

This difference can be fully understood in the context of the
present theory. The detailed calculation addressing this issue was
presented in \cite{06CLP} . The results of the calculation,
presented as the mean velocity profiles for increasing concentration
of the two types of polymers are shown in Figs. \ref{figflex} and
\ref{rod}.
\begin{figure}
\centering \epsfig{width=.40\textwidth,angle=-90,file=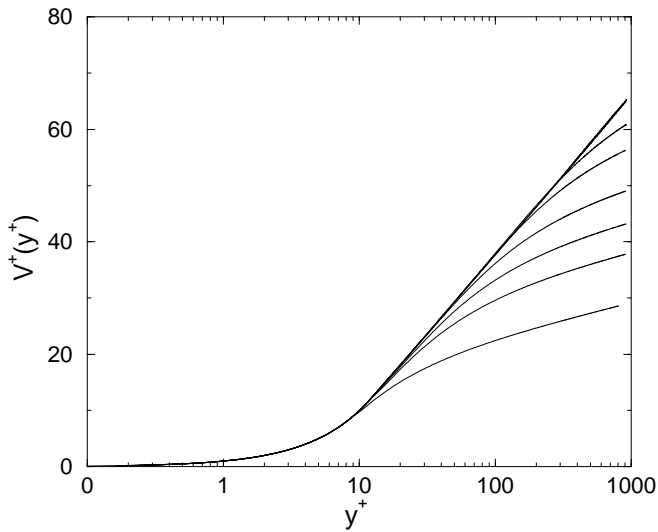}
\caption{The mean velocity profiles for flexible polymer additives
with $\tilde \nu=1,5,10,20,50, 100$ and 500 from below to above.
Note that the profile follows the MDR until it crosses over
back to the Newtonian plug.}
\label{figflex}
\end{figure}
\begin{figure}
\centering \epsfig{width=.40\textwidth,angle=-90,file=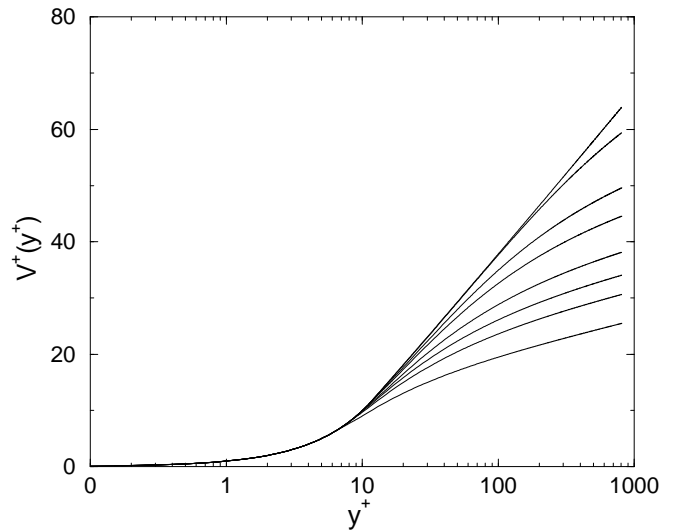}
\caption{The mean velocity profile for rod-like polymer additives
with $\tilde \nu=1,5,10,20,50, 100, 500, 1000, 5000$ and 10000 from below to above.
Note the typical behavior expected for rod-like polyemrs,  i.e. that the profile
diverges from the von K\'arm\'an log-law, reaching the MDR only asymptotically.}
\label{rod}
\end{figure}
The reader should note the difference between these profiles as a
function of the polymer concentration. While the flexible polymer
case exhibits the feature \cite{Virk,97VSW} that the velocity
profile adheres to the MDR until a cross-over to the Newtonian plug
is realized, the rod-like case presents a ``fan" of profiles which
only asymptotically reach the MDR. We also notice that the flexible
polymer matches the MDR faithfully for relatively low values of
$\tilde \nu\equiv \nu_p/\nu_0$, whereas the rod-like case attains the MDR only for
much higher values of $\tilde \nu$. This result is in agreement with
the experimental finding in \cite{Bonn, Bonn2} that the flexible
polymer is a better drag reducer than the rod-like analogue.

The calculation of \cite{06CLP} allows a parameter-free estimate of
the cross-over points $y^+_v$ from the MDR to the ``newtonian plug"
in the case of flexible polymers. The resulting estimate reads
\begin{equation}
y^+_{_{\rm V}} =12+ \tilde \nu \sqrt{0.1} \ .
\end{equation}
These estimates agrees well with the numerical results in Fig.
\ref{figflex}. No such simple calculation is available for the case
of the rod-like polymers since there is no clear point of departure
for small $\tilde \nu$.

We should note that the higher efficacy of flexible polymers
cannot be easily related to their elongational viscosity as
measured in laminar flows. In some studies \cite{Bonn, Bonn2,
Dentoonder} it was proposed that there is a correlation between
the elongational viscosity measured in laminar flows and the drag
reduction measured in turbulent flows. We find here that flexible
polymers do better in turbulent flows due to their contribution to
the effective shear viscosity, and their improved capability in
drag reduction stems simply from their ability to stretch,
something that rod-like polymers cannot do.

\subsection{Cross-over phenomena as a function of the Reynolds number}

Finally, we address the drag enhancement by rod-like polymers when
the values of \Re~ are too small, see Fig. \ref{Virk}. The strategy
of \textcite{Bonn1} is to develop an approximate formula for the
effective viscosity in the case of rod-like polymers that
interpolates properly between low and high values of \Re. For that
purpose we remind the reader of the high \Re~ form of $ \nu_{\rm
eff}$, derive the form for low value of \Re, and then offer an
interpolation formula.

Even for intermediate values of \Re~ one cannot neglect $y$ in the
production term $p'(L-y)$ in the momentum balance equation. Keeping
this term the momentum balance equation becomes, in wall units,
\begin{equation}
\label{mom} (1 + \nu_p^+ R_{yy}) S^+ + W^+ = 1-\frac{y^+}{\Re}\ ,
\end{equation}
where $\nu_p^+ \equiv \nu_p/\nu_0$.

The energy balance equation remains as before,
\begin{equation}
\label{energy1} W^+  S^+  \sim  a^2 (1 + \nu_p^+ R_{yy})
\frac{K^+}{(y^+)^2} +  b \frac{(K^+)^{3/2}}{y^+} \ .
\end{equation}
As was explained above, equations  (\ref{mom}) and (\ref{energy1}) imply that the
polymers change the properties of the flows by replacing the
viscosity by
\begin{equation}
\label{nu} \nu_{\rm eff} = 1 + \nu_p^+ R_{yy} \ .
\end{equation}
In the fully developed turbulent flow with rod-like polymers, when
\Re~ is very large, it was shown in \cite{05BCLLP} that $R_{yy}$
depends on $K^+$ and $S^+$:
\begin{equation}
\label{turbulent} R_{yy} = \frac{K^+}{(y^+ S^+)^2} \ .
\end{equation}
It was argued in \cite{05BCLLP} that for large Re, $K^+$ grows
linearly with $y^+$ and thus the viscosity profile is linear.

Next consider low \Re~flows. According to Eq. (\ref{nu}), the value
of $\nu_{\rm eff}$ depends on $\nu^+_p$ and $R_{yy}$. The value of
$\nu^+_p$ is determined by the polymer properties such as the number
of monomers, their concentration etc., and thus $\nu^+_p$ should be
considered as an external parameter in the equation. The value of
$R_{yy}$, on the other hand, depends on the properties of the flow.
In the case of laminar flow with a constant shear rate, i.e.,
$K^+=W^+=0$ and $S^+= $constant, it was shown theoretically in
\cite{88DE} that:
\begin{equation}
\label{laminar} R_{yy} =  \frac{2^{1/3}}{\De^{2/3}} \ ,
\end{equation}
Thus, the effective viscosity is reduced if $S$ is increased, and
therefore the rod-like polymers solution is a shear-thinning
liquid. Naturally, The value of \De~changes with \Re. To clarify this
dependence we consider the momentum equations Eq.(\ref{finalmom})
at $y=0$ in the Newtonian case.
\begin{equation}
\label{mumbo} \nu_0 S = p'L \ .
\end{equation}
Usually in experiments the system size and the
working fluid remain the same.  Therefore, $\nu_0$ and $L$ are
constants and so \Re~depends on $p'L$ only. According to
(\ref{red}), \Re$_\tau$ grows as $\sqrt{p'L}$ and therefore
\begin{equation}
\label{De} \De = \frac{\nu_0}{\gamma L^2} \Re^2
\end{equation}
 by Eq.(\ref{mumbo}).  Putting into Eq.(\ref{laminar}), we have
\begin{equation}
\label{nu_vis} \nu_{\rm eff}= 1+ \nu^+_p \frac{\lambda}{{
\Re}_{\tau}^{4/3}} \  ,
\end{equation}
where $\lambda \equiv \nu_0/\gamma L^2 $ is a constant.

In the case of intermediate \Re, we need an interpolation between
Eqs. (\ref{turbulent}) and (\ref{laminar}).  To do this we note that
when $y^+$ is small, the solution of Eqs. (\ref{WK}), (\ref{mom})
and (\ref{energy1}) result in $W^+=K^+=0$ in the viscous sub-layer.
This implies that the flow cannot be highly turbulent in the viscous
sub-layer. Thus, it is reasonable to employ Eq.(\ref{laminar}) as
long as $y^+$ is small.
 On the other hand, as the upper bound of $y^+$
is \Re, when $y^+$ is large, it automatically implies
that \Re~is large. The laminar contribution is therefore
negligible as it varies inversely with \Re. The effective
viscosity due to the polymer is dominated by the turbulent
estimate, Eq. (\ref{turbulent}). To connect these two regions we
simply use the pseudo-sum:
\begin{eqnarray}
 \nu_{\rm eff} &=& 1+ \nu^+_p \left(
\frac{\lambda}{\Re^{4/3}} + \frac{K^+}{(y^+ S^+)^2}
 \right) \nonumber \\ &=& g+ \nu^+_p \frac{K^+}{(y^+ S^+)^2} \label{nuall} \  ,
\end{eqnarray}
where $g\equiv 1+ \nu^+_p \lambda /{\Re}^{4/3}$. One can
see that the limits for both high and low  \Re are satisfied.

The form for (\ref{nuall}) for $\nu_{\rm eff}$ was incorporated into
the balance equations which were analyzed and solved
self-consistently in \cite{Bonn1} one of the most interesting
results, which can be directly compared to the data in Fig.
\ref{Virk} refers tot the drag reduction as a function of the
concentration of the rod-like polymers, presented in
Prandtl-K\'arm\'an coordinates. The results of the theoretical
predictions are shown in Fig. \ref{Redep}, and the qualitative
agreement with the experimental observations is obvious.
\begin{figure}
  \includegraphics[width=3.5in]{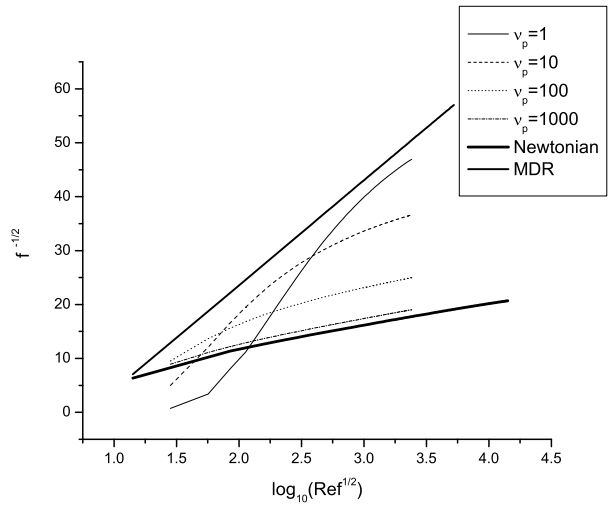}\\
  \caption{$f^{-1/2}$ as the function of $\log_{10}(\Re f^{1/2})$ with
$\lambda=1$ with various values of
$\nu_p$.}\label{Redep}
\end{figure}
Another interesting comparison with experimental findings is
available due the work presented in \cite{Bonn1} where the
percentage of drag enhancement and reduction were measure as a
function of $\nu_p$. The quantitative comparison needs a careful
identification of the material parameters in the theory and the
experiment, and this was described in detail in \cite{Bonn1}. Fig.
\ref{DR1} shows the comparison of the percentage of drag reduction
(enhancement) between the theoretical predictions and the
experimental results. The two data sets shown pertain to
$c_p=250$wppm and $c_p=500$wppm. The agreement between theory and
experiment is very satisfactory.
\begin{figure}
  \includegraphics[width=3.5in]{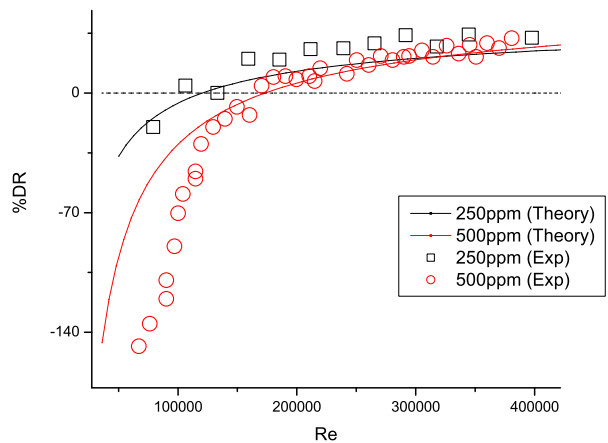}\\
  \caption{Percentage drag reduction as a function of Re for two values of the concentration of
  the rod-like polymer. Symbols are results of experiments and solid lines--theory, cf.  \cite{Bonn1} }\label{DR1}
\end{figure}

\section{Drag Reduction by Bubbles}

In order not to leave the reader with the impression that polymers
are the only additives that can reduce the drag, or that they
provide the only technologically preferred method, we discuss
briefly drag reduction by other additives like surfactants and
bubbles \cite{Book1}. Generally speaking, the understanding of drag
reduction by these additives lags behind what had been achieved for
polymers. The importance of drag reduction by bubbles cannot be
however overestimated;  for practical applications in the shipping
industry the use of polymers is out of the question for economic and
environmental reasons, but air bubbles are potentially very
attractive \cite{jap}.

The theory of drag reduction by small concentrations of minute
bubbles is relatively straightforward, since under such conditions
the bubbles only renormalize the density and the viscosity of the
fluid, and a one-fluid model suffices to describe the dynamics
\cite{Itamar1}. The fluid remains incompressible, and the equations
of motion are basically the same as for a Newtonian fluid with
renormalized properties. The amount of drag reduction under such
conditions is however limited. But when the bubbles increase in
size, the one-fluid model loses its validity since the bubbles
become dynamical in the sense that they are no longer Lagrangian
particles, their velocity is no longer the fluid velocity at their
center, and they begin to fluctuate under the influence of local
pressure variations. The fluctuations of the bubbles are of two
types: 1) the bubbles are no longer spherical, distorting their
shape according to the pressure variations, and 2) the bubbles can
oscillate {\em radially} (keeping their spherical shape) due to the
compressibility of the gas inside the bubble.  The first effect was
studied numerically using the ``front tracking" algorithm of
\textcite{Jap3, Tryggvason}.  However, the results indicate either a
drag enhancement, or a limited and transient drag reduction.  This
leads one to study the possibility of explaining bubbly drag
reduction by bubble oscillations.  Indeed, a theoretical model
proposed by \textcite{Legner} successfully explained the bubbly drag
reduction by relating turbulent viscosity in the bubbly flow to the
bulk viscosity of the bubbles. While the bulk viscosity is important
only when the bubbles are compressible, it is important and
interesting to see how and why it affects the charactistics of the
flow. Here we review how drag reduction is caused by bubbles when
bubble oscillations are dominant.  Finally we compare our finding
with the results of \textcite{Legner}, showing that a nonphysical
aspect of that theory is removed, while a good agreement with
experiment is retained.

\subsection{Average equations for bubbly flows: the additional stress tensor}
\label{aveq} Dealing with bubbles we cannot put the fluid density to unity any longer and we
must display it explicitly. Here a Newtonian fluid with density $\rho$ is laden with
bubbles of density $\rho\Sub B$, and radius $R$ which is much
smaller than the outer scale of turbulence $\C L$. The volume
fraction of bubbles $C$ is taken sufficiently small such that the
direct interactions between bubbles can be neglected. In writing the
governing equations for the bubbly flow we will assume that the
length scales of interest are larger than the bubble radius. Later
we will distinguish however between the case of microbubbles whose
radius is smaller than the Kolmogorov scale $\eta$ and bubbles whose
radius is of the order of $\eta$ or slightly larger. For length
scales larger than the bubbles one
writes~\cite{stress1,stress2,stress3}:
\begin{equation}
\label{basic}
\rho (1-C) [ \frac{\partial {\B U }  }{\partial t} +\B U \cdot \B\nabla \B U]=
- \B \nabla  p  +\B \nabla \cdot {\B  \sigma}  +C\, \B\nabla \cdot {\B \tau}\ .
\end{equation}
\begin{equation}
\frac{\partial (1-C)}{\partial t} + \B \nabla \cdot (1-C) \,{\B U} =0 \ .
\end{equation}
In these equations, $\B U$ is the velocity of the carrier fluid, and
\begin{equation}
\sigma_{ij} \equiv \frac{\mu_0}{2} \(\frac{\partial U_i}{\partial x_j}
+\frac{\partial U_j}{\partial x_i}\) \ .
\end{equation}
Here $\mu_0=\rho \nu_0$. The effect of the bubbles appears in two
ways, one through the normalization of the fluid density, $\rho\to
\rho (1-C)$, and the second is via the extra stress tensor ${\B
\tau}$. Note that in the case of polymers we accounted only for the
additional stress tensor  since the actual volume fraction of
polymers is minute, typically less then $10^{-3}$. A typical value
of the bubbles volume fraction is 0.05 or even more, and cannot be
neglected.

The evaluation of the extra stress tensor $\B \tau$ is tedious.
There are a number of forces acting on bubbles, but following
\textcite{06LLP} we neglect gravity and the lift force. In
\cite{06LLP} it was found that the add-mass force due to bubble
oscillations and the add-mass force due to bubble acceleration are
of the same order of magnitude. The viscous forces were of course
taken into account. The result is of the following form:
\begin{eqnarray}\nonumber
{\B \tau}  &=& \rho \Big\{\Big[  \frac14\,({\B  w}- \B U)
 \cdot ({\B  w}- \B U) - R  \ddot{R}-\frac52\, \dot{R} ^2 \Big]
{\B  I} \\
\label{tau1} &&\quad-
\frac12\,(
{\B  w}- \B U) ( {\B  w} - \B U)+\frac12\, \mu_0 \B \sigma \Big\}\,,
\end{eqnarray}
Here $\B I$ is the unit tensor, $\B  w$ is the bubble velocity which differs from the carrier
velocity $\B U$. $\dot R$ stands for the rate of change of the
bubble radius $R$.  The relative importance of the various terms in
$\B \tau$ depends on the values of Re and also on the so called
Weber number,  which is defined as
\begin{equation}
{\rm We} \equiv \frac{\rho |{\B  w}- \B U|^2 R}{\gamma} \ .
\end{equation}
The Weber number is the ratio of the kinetic energy of the fluid
associated with the bubble motion over the surface energy of the
bubble itself due to surface tension $\gamma$. When We$\ll 1$ the
bubbles can be considered as rigid spheres. If We is sufficiently
large,  the bubbles begin to deform and oscillate; this contributes
a significant contribution to ${\B \tau }$. In the following section
we show that this can be crucial for drag reduction.

For very small bubbles (micro-bubbles) of very small density the
last term in $\B \tau$ , i.e. $ \mu_0 \B S$ which is the viscous
contribution, is the only one that survives.  When this is the case
the bubble contribution to the stress tensor can be combined with
$\B \sigma$ in  Eq, (\ref{basic}),  resulting in the effective
viscosity given by
\begin{equation}\label{muef}
 \mu_{\rm eff} =
\big(1+\frac{5}{2} C\big) \mu\ .
\end{equation}
The study of drag reduction under this renormalization of the
viscosity and the density was presented in \cite{Itamar1} with the
result that drag reduction can be obtained by putting the bubbles
out of the viscous sub-layer and not too far from the wall. The
amount of drag reduction is however rather limited in such
circumstances.
\subsection{Balance equations in the turbulent boundary layer}
\label{bal} At this point we apply the formalism detailed above to
the question of drag reduction by bubbles in a stationary turbulent
boundary layer with plain geometry. This can be a pressure driven
turbulent channel flow or a plain Couette  flow, which is close to
the circular Couette flow realized in ~\cite{Detlef}. As before , we
take the smallest geometric scale to be $2L$, the unit vector in the
streamwise and spanwise directions be $\hat x$ and $\hat z$
respectively, and the distance to the nearest wall be $y\ll L$. The
variables that enter the analysis are modified by the density,
except for  the mean shear, which remains as in Eq. (\ref{SWK}). The
turbulent kinetic energy density is modified to read
\begin{equation}\label{KE}
{\C K}\=\frac12 \rho(1-C)\langle |\B u|^2\rangle \ ,
\end{equation}
and the  Reynolds stress 
\begin{equation}\label{RS}
{\C W}\=-\rho(1-C)\langle u_xu_y \rangle\ .
\end{equation}
\subsubsection{\label{ss:bal-m}Momentum balance}

From Eq.  (\ref{basic}) we derive the exact equation for the
momentum balance by averaging and integrating in the usual way, and
find for $y\ll L$:
\begin{equation}
\label{mombal}
\rho P = \mu_0 S +  \C W+ C \langle \tau_{xy} \rangle \ .
\end{equation}
Here $P$ is the momentum flux toward the wall.

For $C=0$ Eq. (\ref{mombal}) is the usual Eq. (\ref{finalmom})
satisfied by Newtonian fluids. To expose the consequences of the
bubbles we notice that the diagonal part of the bubble stress tensor
$\B \tau$ [the first line in the RHS of Eq. (\ref{tau1})] does not
contribute to Eq. (\ref{mombal}). The $xy$ component of the
off-diagonal part of $\B \tau$ is given by the 2nd line in Eq.
(\ref{tau1}). Define the dimensionless ratio
\begin{equation}
\zeta \equiv \frac{\langle(w_x- U_x)(w_y- U_y)\rangle}
 {2 \langle u_xu_y \rangle} \  . \label{defalf}
\end{equation}
For later purposes it is important to assess the size and sign of
$\zeta$. This object was analyzed in \cite{06LLP}, and it was shown
that for small values of Re, $\zeta$ is small. On the other hand,
for large Re the fluctuating part of $\B w$ is closely related to
the fluctuating part of $\B u$. The relation is
\begin{equation}
\B w -\B U\approx 2\B u \ . \label{estimate}
\end{equation}
This implies that
$\zeta\approx 2 $ and is positive definite, as we indeed assume bellow.
With this definition we can simplify the appearance of Eq.
(\ref{mombal}):
\begin{equation}
\rho P = \mu_{\rm eff} S + \frac{1+C\(\zeta-1\) }{1-C}\C W \,,\label{finalmomB}
\end{equation}
with $\mu_{\rm eff}$ defined by (\ref{muef}).
\subsubsection{\label{ss:bal-e}Energy balance}
Next, we consider the balance of turbulent energy in the log-layer.
In this region, the production and dissipation of turbulent kinetic
energy is almost balanced.  The production can be calculated
exactly, $\C W S$. The dissipation of the turbulent energy is
modeled by the energy flux which is the kinetic energy $\C K(y)$
divided by the typical eddy turn over time at a distance $y$ from
the wall, which is $\sqrt{\rho(1-C)}y/b\sqrt{\C K}$ where $b$ is a
dimensionless number of the order of unity. Thus the flux is written
as $b \C K^{3/2}\big /y \sqrt{\rho (1-C)}$. The extra dissipation
due to the bubble is $ C \langle \tau_{ij}s_{ij} \rangle$ where
$s_{ij}\equiv \partial u_i/\partial x_j$. In summary, the turbulent
energy balance equation is then written as:
\begin{equation}
\label{energy}
 \frac{b\,  \C K^{3/2}}{\sqrt{\rho (1-C)}y} +
 C \langle \tau_{ij}s_{ij} \rangle =  \C W S\ .
\end{equation}
As usual, the energy and momentum balance equations do not close the
problem, and we need an additional relation between the objects of
the theory, and we define $c\Sub B$ in a manner following Eq.
(\ref{WK}):
\begin{equation}
\label{WKBa}
\C W \equiv c\Sub B^2 \C K \ .
\end{equation}
Clearly, $\displaystyle \lim_{C\to 0}c\Sub B=c\Sub N$ and for small
$C$ (noninteracting bubbles)  $c\Sub B^2-c\Sub N^2\propto C$. Having
in mind that in numerical simulations of the incompressible bubbly
flow, it was reported~\cite{Chahed, Lance} that $c\Sub B$ is
slightly smaller than its Newtonian counterpart we can write
\begin{equation}\label{WKBab}
c\Sub B^2=c\Sub N^2(1- \tilde\beta\,C)\,,
\end{equation}
with positive coefficient $\tilde\beta$ of the order of unity. We
are not aware of direct measurements of this form in bubbly flows,
but it appears natural to assume that the parameter $\tilde\beta$ is
$y$-independent in the turbulent log-law region.
\subsection{Drag reduction in bubbly flows}
\label{drag}

In this section we argue that  bubble oscillations are crucial in
enhancing the effect of drag reduction. This conclusion is in line
with the experimental observation of \cite{Detlef} where bubbles and
glass spheres were used under similar experimental conditions.
Evidently,  bubble deformations can lead to the compressibility of
the bubbly mixture. This is in agreement with the simulation of
\cite{Said} where a strong correlation between compressibility and
drag reduction were found.

To make the point clear we start with the analysis of the energy
balance equation (\ref{energy}). The additional stress tensor
$\tau_{ij}$ Eq. (\ref{tau1}) has a diagonal and an off diagonal
part. The off-diagonal part has a viscous part that is negligible
for high Re. The other term can be evaluated using the estimate
(\ref{estimate}), leading to a contribution to $\langle \tau_{ij}
s_{ij}\rangle$ written as
\begin{equation}
\langle \frac{1}{2}( {\B  w}- \B U) ( {\B  w} - \B U)
 : \B \nabla \B u\rangle \approx 2 \langle  \B u \B u :
  \B \nabla \B u\rangle \ . \label{Uu}
\end{equation}
The expression on the RHS is nothing but the spatial turbulent
energy flux which is known to be very small  in the log-layer
compared to the production term on the RHS of Eq. (\ref{energy}). We
will therefore neglect the off-diagonal part of the stress tensor in
the energy equation. The analysis of the diagonal part of the stress
depends on the issue of bubble oscillations and we therefore discuss
separately oscillating bubbles and rigid spheres.
\subsubsection{Drag reduction with rigid spheres}
Consider first situations in which $\dot{R}=0$.  This is the case
for bubbles at small We, or when the bubbles are replaced by some
particles which are less dense than the carrier fluid \cite{Detlef}. When
the volume of the bubbles is fixed, the incompressibility condition
for the Newtonian fluid is unchanged, and $s_{ii}=0$. The diagonal
part of  ${\bm \tau}$, Eq. (\ref{tau1}), due to the
incompressibility condition $s_{ii}=0$, has no contribution to $
\langle \tau_{ij}s_{ij} \rangle$.  Due to the discussion after Eq.
(\ref{Uu})  the term due to the extra stress tensor can be neglected
and the energy balance equation is then unchanged compared to the
Newtonian fluid. Note that the momentum balance equation is
nevertheless affected by the bubbles.  Putting (\ref{WKBa}) into
(\ref{energy}), we have after simple algebra
\begin{equation}
\C W=\rho(1-C) \frac{S^2 y^2 c\Sub B^6 }{b^2}\ . \label{amazing}
\end{equation}
To assess the amount of drag reduction we will consider an
experiment \cite{jap} in which the velocity profile (and thus $S$)
is maintained fixed. Drag reduction is then measured by the
reduction in the momentum flux $P$. Substituting Eq. (\ref{amazing})
into Eq. (\ref{finalmomB}) in which one can neglect $\mu_{\rm eff}
S$, and replacing $S(y) y =1/\kappa_{_{\rm K}}$ we find
\begin{equation}
P=\frac{ \rho (1-C+\zeta C) c\Sub B^6 }{\kappa_{_{\rm K}}^2 b^2}.
\end{equation}
If there are no bubbles ($C=0$), the Newtonian momentum flux $P\Sub N$ reads
\begin{equation}
P\Sub N = \frac{\rho\,  c\Sub N^6}{\kappa_{_{\rm K}}^2 b^2} \ . \label{Prigid}
\end{equation}
The percentage of drag reduction can be defined as
\begin{eqnarray}\nonumber
{\%}\mbox{DR}&=&\frac{P\Sub N-P}{P\Sub N}=1-
\frac{(1-C+\zeta C) c\Sub B^6}{c\Sub N^6}\\ \label{DR}
&\approx &\(1-\zeta+3\tilde\beta \) C\ .
\end{eqnarray}
At small Re, $\zeta=0$ and the amount of drag reduction increases
linearly with $C$.  If Re is very large we expect $\zeta\approx 2$,
and then the drag is {\em enhanced}.  This result is in pleasing
agreement with the experimental data in \cite{Detlef}.  Indeed, the
addition of glass beads with density less than water caused drag
{\em reduction} when Re~ is small, whereas at Re~$ \sim (10^6)$, the
drag was slightly {\em enhanced}.

\subsubsection{Drag reduction with flexible bubbles}
If the value of We is sufficiently large such that $\dot{R} \neq 0$,
the velocity field is no longer divergenceless.  Also, to evaluate
the extra stress tensor one needs the explicit dynamical equation
for the bubble radius. This equation was provided by
\textcite{stress1}, and was analyzed carefully in \cite{06LLP}.
Using the results of the analysis in Eq. (\ref{tau1}) one can
simplify the extra stress tensor (for calculating the correlation
$\langle \tau_{ij} s_{ij}\rangle)$) to the form
\begin{equation}
{\bm\tau} \approx  \rho \left[ \frac{4}{3}\mu \nabla
\cdot u {\B  I} + \frac{1}{2}(\B  w-\B U)(\B  w-\B U) \right]
\end{equation}
The extra turbulent dissipation due to the bubble is
 $\langle \tau_{ij}s_{ij} \rangle$. In light
of the smallness of the term in Eq. (\ref{Uu}) we find
\begin{eqnarray}
\langle \tau_{ij}s_{ij} \rangle &=& \langle  \frac{4}{3} \mu s_{ii}^2   \rangle \ .
\end{eqnarray}
The term $ \frac{4}{3}\mu s_{ii}^2$ is of the same form as the usual
dissipation term $\mu s_{ij}s_{ij}$ and therefore we write this as:
\begin{equation}
\langle \frac{4}{3}\mu s_{ii}^2 \rangle = A \frac{\rho |{\B u}|^{3/2}}{y}=
 A \frac{\C K^{3/2}}{\sqrt{\rho} (1-C)^{3/2} y}
\end{equation}
where $A$ is an empirical constant.  Finally, the energy equation becomes
\begin{equation}
\label{en2}
\frac{ b (1-C) +A C }{\rho \sqrt{1-C}}\frac{\C K^{3/2}}{y} = \C W S
\end{equation}
\begin{figure}
\centering\epsfig{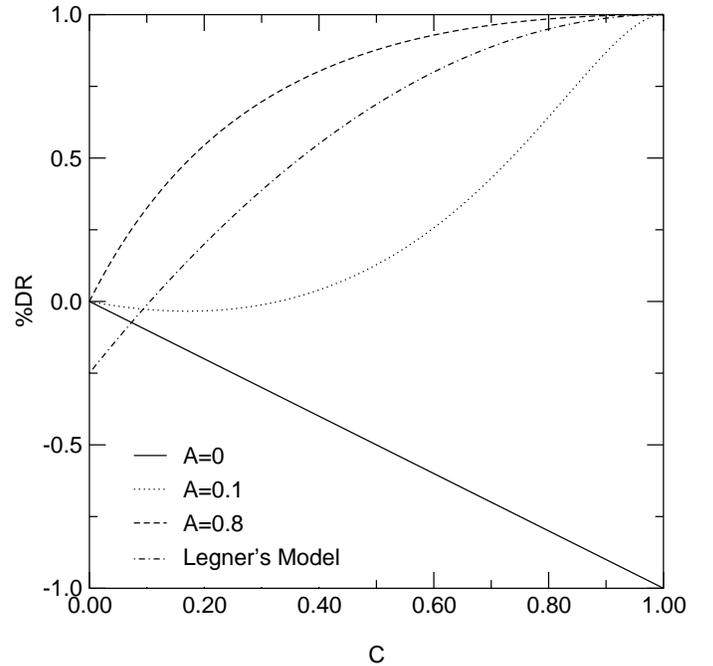}
\caption{Predicted values of drag reduction with $\alpha=2$ and
different values of $A$. In a dashed line we reproduce the
predictions of Legner's theory which suffer from an unphysical drag
enhancement at $C=0$, For $A=0$ (rigid spheres) we find only
increasing drag enhancement as a function of $C$. For small values
of $A$ we have first a slight drag enhancement, and then modest drag
reduction. For large values of $A$, associated with strong bubble
oscillations, we find significant values of drag reduction.}
\label{bubbles}
\end{figure}
As we derived Eq. (\ref{Prigid}), we again specialize the situation
to an experiment in which $S$ is constant, and compute the momentum
flux
\begin{equation}
P= \frac{ (1-C)^2(1-C+\zeta C)}{(1- C + \frac{A}{b} C)^2}
\frac{c\Sub B^6}{\kappa_{_{\rm K}}^2 b^2} \ . \label{Pflex}
\end{equation}
The degree of drag reduction is then
\begin{equation}
\% DR = 1 - \frac{(1-C)^2 (1-C+ \zeta C)}{(1-C +\frac{A}{b} C)^2}
(\frac{c\Sub B}{c\Sub N})^6 \ .
\end{equation}
Note that $A$ is an unknown parameter that should depend on $We$,
and so its value is  different in different experiments. The
percentage of drag reduction for various values of $A$ are shown in
Fig. \ref{bubbles} where we chose $\alpha=2$ and for simplicity we
estimate $c\Sub B=c\Sub N$. One sees that for $\alpha=2$ and $A=0$
(where the latter is associated with rigid bubbles), we only find
drag enhancement. For small value of $A$, or small amplitudes of
oscillations, small concentrations of bubbles lead (for $\alpha=2$)
to drag enhancement, but upon increasing the concentration we find
modest drag reduction.   Larger values of $A$ lead to considerably
large degrees of drag reduction.  For $A=0.15$, the result agrees
reasonably with Legner's theory which predicts $\% DR \approx
1-5(1-C)^2/4$ \cite{Legner}. Note that according to Legner, there
should be considerable drag enhancement when $C=0$.  This is of
course a nonsensical result that is absent in our theory.  For
$A=0.8$, $\% DR \approx 4 C$ for small $C$. This is the best fit to
the experimental results which are reported in \cite{jap}.
\section{Summary and Discussion}
\label{summary}

We offered a review of drag reduction by polymers and bubbles. Quite
generally, we have shown that drag reduction can be understood in
great detail by using only a few equations that govern the budgets
of energy and momentum. Both polymers and bubbles open up an
additional channel for dissipation, and thus pose a fundamental
riddle: why do they reduce the drag. The answer in all cases is
fundamentally the same: the same agents reduce the momentum
flux from the bulk to the wall, and this is the main effect leading
to drag reduction. The reduction in the momentum flux overwhelms the
increase dissipation. We stressed above, and we re-iterate here,
that this mechanism depends on the existence of a wall which breaks
the translational symmetry. Drag reduction must be discussed in the
context of wall-bounded flows to make full sense. Of course,
additives may influence also the spectrum of turbulent fluctuations
in homogeneous and isotropic turbulence, but this is another story,
quite independent of drag reduction.

One should make a great distinction between the two cases. The
phenomenon of drag reduction by polymers exhibits interesting
universal features which are shared even by flexible and rod-like
polymers. These features are the most prominent experimental results
that required theoretical understanding. We have shown that the MDR
has a special significance in being an edge solution of turbulent
flows. Trying to increase the reduction of the drag behind what is
afforded by the MDR would re-laminarize the flow. This may be the
central theoretical insight that is offered in this review,
providing a simple and intuitive meaning to the nature of the MDR.
This explains why flexible and rod-like polymers have the same MDR,
even though they approach the MDR in distinctly different ways.

Once the theory was put forth to explain the universal aspects of
drag reduction by polymers, it became also clear that it can easily
explain, in considerable quantitative detail, also the non-universal
aspects, including cross-overs due to small concentrations of
polymers, low values of \Re, and small number of monomers in the
polymer chains. We trust that this detailed understanding can help
in designing and optimizing the use of polymers in practical
applications of drag reduction.

On the other hand, the case of bubbles exhibits much less
universality, since the placement of the bubbles with respect to the
wall and their actual density profile have crucial consequences
regarding their efficacy as drag reducers. The main conclusion of
our study is that bubble oscillations can contribute decisively to
drag reduction in turbulent flows. In agreement with the
experimental findings of \cite{Detlef}, we find that rigid bubbles
tend to drag enhance, and the introduction of oscillations whose
amplitude is measured by the parameter $A$ (Fig. \ref{bubbles})
increases the efficacy of drag reduction.  The main drawback of the
present study is that the bubble concentration was taken uniform in
the flow. In reality a profile of bubble concentration may lead to
even stronger drag reduction if placed correctly with respect to the
wall. A consistent study of this possibility calls for the
consideration of buoyancy and the self-consistent solution of the
bubble concentration profile. Such an effort is beyond the scope of
this review and must await future progress.

\begin{acknowledgments}
We thank our collaborators Emily S.C. Ching , Elisabetta DeAngelis,
Ting-Shek Lo, Anna Pomyalov and Vasyl Tiberkevich without whom the
research reviewed in this paper could not be accomplished. Albert Libchaber and Ting-Shek
Lo have read the manuscript and made a number of valuable comments
and suggestions.  This work has been supported in part by the
US-Israel Binational Science Foundation, by the European Commission
under a TMR grant, and by the Minerva Foundation, Munich, Germany.
\end{acknowledgments}

\vskip 0.2 cm
\appendix*
\section{The Hidden Symmetry of the Balance Equations}
Consider the following identity:
\begin{eqnarray}
\nu^+(y^+)&=&1+\alpha(y^+-\delta^+)\nonumber\\ &=&[1+\alpha
(y^+-\delta)+\alpha ( \delta-\delta^+)]\nonumber\\
&=&g(\delta) \left[1+\frac{\alpha}{g(\delta)} (y^+ -  \delta)\right] \ ,
\end{eqnarray}
where
\begin{equation}
g( \delta)\equiv 1+\alpha( \delta - \delta^+)\ , \quad
 \delta\ge \delta^+ \ . \label{gt}
\end{equation}
Next introduce newly renormalized units using the effective
viscosity $g( \delta)$, i.e.
\begin{equation}
y^\ddag\equiv \case{y^+}{g( \delta)} , \quad \delta^\ddag
\equiv \case{\delta}{g( \delta)} \ , \quad S^\ddag \equiv S^+ g( \delta) \ ,
\quad W^\ddag \equiv W^+  \ . \label{vispr}
\end{equation}
In terms of these variables the balance equations are rewritten as
\begin{eqnarray}
&&[1+\alpha(y^\ddag-\delta^\ddag)]S^\ddag+ W^\ddag =1\ , \label{vepr1}\\
&&[1+\alpha(y^\ddag-\delta^\ddag)]\frac{{\Delta}^2(\alpha)}{{y^\ddag}^2}
+\frac{\sqrt{W^\ddag}}{\kappa_{_{\rm K}} y^\ddag} =S^\ddag \ . \label{vepr2}
\end{eqnarray}
These equations are isomorphic to (\ref{pr1}) and (\ref{pr2}) with
$\delta^+$ replaced by $\delta^\ddag$. The ansatz (\ref{scaling}) is
then replaced by $\Delta(\alpha) =\delta^+ g( \delta)^{-1} f(\alpha
\delta^\ddag)$ This form is dictated by the following
considerations: (i) $\Delta(\alpha)\to \delta^+$ when $\alpha\to 0$,
(ii) all lengths scales in the re-scaled units are divided by $g(
\delta)$, and thus the pre-factor in front of $f$ becomes
$\delta^+/g( \delta)$, and (iii) $\alpha \delta^+$ in Eq.
(\ref{bal2}) is now replaced in Eq. (\ref{vepr2}) by
$\alpha\delta^\ddag$, leading to the new argument of $f$. Since the
function $\Delta(\alpha)$ cannot change due to the change of
variables, the function $\Delta(\alpha)$ should be identical to that
given by Eq. (\ref{scaling}):
\begin{equation}
\delta^+f(\alpha\delta^+)=\frac{\delta^+}{g( \delta)} f(\alpha \delta^\ddag) \ .
\end{equation}
Using the explicit form of $g( \delta)$ Eq. (\ref{gt}), and choosing
(formally first) $\delta=\delta ^\ddag=0$ we find that
$f(\xi)=1/(1-\xi)$. It is easy to verify that this is indeed the
solution of the above equation for any value of $\delta^\ddag$, and
therefore the unique form of Eq. (\ref{Delta}).



\begin{thebibliography}{99}
\bibitem[Amarouchene et al.(2007)Amarouchene et al.]{Bonn1}
Y. Amarouchene, D. Bonn, H.Kellay, T.S. Lo, V. S.
 L'vov and I Procaccia, 2007, ``Reynolds number dependence of drag
reduction by rod-like polymers", submitted to Phys. Fluids, Also:
nlin.CD/0607006

\bibitem[Bellakhel et. al.(2004)Bellakhel et al]{Chahed}
G. Bellakhel, J. Chahed and L. Masbernat, 2004, J. Turbul. {\bf 5}, 036.

\bibitem[Benzi et al.(2004a)Benzi et al.]{04BCHP}
R. Benzi, E. Ching, N. Horesh and I. Procaccia, 2004,Phys. Rev.
Lett. {\bf 93}, 078302.

\bibitem[Benzi et al.(2004b)Benzi et al.]{04BLPT}
R. Benzi, V. S. L'vov, I. Procaccia and V. Tiberkevich, 2004
Europhys. Lett, {\bf 68},  825.

\bibitem[Benzi et al.(2005a)Benzi et al.]{05BDLP}
R. Benzi, E. deAngelis, V.S. L'vov and I. Procaccia,2005, Phys. Rev.
Lett., {\bf 95} 194502.

\bibitem[Benzi et al.(2005b)Benzi et al.]{05BCLLP}
R. Benzi, E. S.C. Ching, T. S. Lo, V. S. L'vov, and I. Procaccia,
2005, Phys. Rev. E, {\bf 72}, 016305.


\bibitem[Benzi et al.(2006)Benzi et al.]{06BDLPT}
R. Benzi, E. de Angelis, V. S. L'vov, I. Procaccia and V.
Tiberkevich, 2006, J. Fluid Mech.  551, 185.

\bibitem[van den Berg et al.(2005)van den Berg et al.]{Detlef}
T. H. van den Berg, S. Luther, D. P. Lathrop and D. Lohse, 2005,
Phys. Rev. Lett., {\bf 94} 044501.

\bibitem[Beris and Edwards(1994)Beris and Edwards]{94BE}
A.N. Beris and  B.J. Edwards, 1994, \textit{Thermodynamics of
Flowing Systems with Internal Microstructure} (Oxford University
Press).

\bibitem[Biesheuvel and Wijngaarden(1984)Biesheuvel and Wijngaarden]{stress2}
A. Biesheuvel and L. wan Wijngaarden, 1984, J. Fluid Mech. , {\bf 148} 301.


\bibitem[Bird et al(1987)Bird et al.]{87BCAH}
R.B. Bird, C.F. Curtiss, R.C. Armstrong, O. Hassager, 1987,
\textit{Dynamics of Polymeric Fluids} (Wiley, NY).

\bibitem[Bonn et al.(2005)Bonn et al]{Bonn2}
D. Bonn, Y. Amarouch\`ne, C. Wagner, S. Douady and O. Cadot, 2005, J.
Phys.: Condens. Matter, {\bf 17}, S1195.

\bibitem[Brenner(1974)brenner]{74Bre}
H. Brenner, 1974, Int. J. Multiphase Flow {\bf 1}, 195.

\bibitem[Ching et al.(2006)Ching et al.]{06CLP}
E. S.C. Ching, T. S. Lo, I. Procaccia, 2006,  Phys. Rev. E, {\bf 74}, 026301.


\bibitem[Choi et al.(2002)Chi et al.]{02CLLC}
H.J. Choi, S.T. Lim, P-Y Lai and C.K. Chan, 2002, Phys. Rev. Lett, {\bf 89} 088302-1.

\bibitem[De-Angelis et al.(2003)De-Angelis et al.]{03Angel}
E. De Angelis, C. M. Casciola, V. S. L'vov, R.Piva and I. Procaccia, 2003
Phys. Rev. E., {\bf 67} 056312 (2003).

\bibitem[De-Angelis et al.(2004)De-Angelis et al.]{04DCLPPT}
E. De Angelis, C. Casciola, V. S. L'vov, A. Pomyalov, I. Procaccia and V.
Tiberkevich, 2004. Phys. Rev. E, {\bf 70}, 055301.

\bibitem[De-Angelis et al.(2005)De-Angelis et al.]{DCBP}
E. De Angelis, C. M. Casciola, R. Benzi and R.Piva, 2005, J. Fluid Mech. {\bf 531}, 1.

\bibitem[Doi and Edwards(1988)Doi and Edwards]{88DE}
M. Doi and S. F. Edwards, 1988, \textit{The Theory or Polymer
Dynamics}, Oxford University Press.


\bibitem[de-Gennes(1979)de-Gennes]{Gen}
P.-G. de-Gennes, 1979, \textit{Scaling Concepts in Polymer Physics}
Cornell University University.

\bibitem[de-Gennes(1990)de-Gennes]{90Gennes}
P.-G. de-Gennes, 1990, \textit{Introduction to Polymer Dynamcis}
Cambrdige University Press.

\bibitem[Gyr and Bewersdorff(1995)Gyr and Bewersdorff]{Book1}
A. Gyr and H. W. Bewersdorff, 1995, \textit{Drag Reduction of
Turbulent Flows by Additives} Kluwe, London.


\bibitem[Dimitropoulos et al.(1998)Dimitropoulos et al.]{98Dim}
C.D. Dimitropousols, S. Sureshkumar and A.N. Beris, 1998, J.
Non-Newtonian Fluid Mech. {\bf 79}, 433 (1998).


\bibitem[Dimitropoulos et al.(2005)Dimitropoulos et al.]{05Dim}
C.D. Dimitropousols, Y. Dubief, E.S.G. Shawfeh, P. Moin and S.K.
Lele, 2005, Phys. Fluids, {\bf 17}, 011705-1.

\bibitem[Escudier et al.(1999)Escudier et al.]{Escudier}
M. P. Escudier, P. Presti and S. Smith, 1999, J. Non-Newtonian Fluid
Mech. {\bf 81}, 197.

\bibitem[Ferrante and Elgobashi(2004)Ferrante and Elgobashi]{Said}
A. Ferrante and S. Elghobashi, 2004, J. Fluid Mech. , {\bf 503}  345.


\bibitem[Flory(1953)Flory]{Flory}
P.J. Flory, 1953 \textit{Principles of Polymer Chemistry}, (Cornell University)

\bibitem[Hinch and Leal(1975)Hinch and Leal]{75HL}
E. Hinch and L. Leal, 1975, J. Fluid Mech. {\bf 71}, 481.

\bibitem[Hinch and Leal(1976)Hinch and Leal]{76HL}
E. Hinch and L. Leal, 1976, J. Fluid Mech. {\bf 76}, 187.

\bibitem[Hoyt(1972)Hoyt]{72Hoyt}
J.W. Hoyt, 1972, Trans. ASME:J. Basic Engng {\bf 94}, 258.

\bibitem[Kawamura and Kodama(2002)Kawamura and Kodama]{Jap3}
T. Kawamura and Y. Kodama, 2002, Int. J. Heat and Fluid Flow {\bf 23}, 627.

\bibitem[Kitagawa et al.(2005)Kitagawa et al.]{jap}
A. Kitagawa, K. Hishida and Y.  Kodama, 2005 Experiment in Fluids , {\bf 38}  466.


\bibitem[Landhal(1973)Landhal]{73Landhal}
M.T. Landhal, 1973, \textit{in Proc. 13th Intl. Congr. Theor. Appl.
Mech., Moscow} (ed. E. Becker and G.K. Mikhailov), 179. Springer.

\bibitem[Lamb(1879)Lamb]{Lamb}
H. Lamb, 1879, \textit{Hydrodynamics}, Dover reprint 1945.

\bibitem[Lance et. al. (1991)Lance et. al]{Lance}
M. Lance, J.L. Marie and J. Bataille, 1191, J. Fluid Eng. {\bf 113}, 295.

\bibitem[Legner(1984)Legner]{Legner}
H. H. Legner, 1984, Phys. Fluids , {\bf 27}  2788.


\bibitem[Lumley(1969)Lumley]{69Lumley}
J.L. Lumley, 1969, Ann. Rev. Fluid Mech. {\bf 1}, 367.

\bibitem[Lo et al.(2005)Lo et al.]{05LLPP}
T.S. Lo, V. S. L'vov, A. Pomyalov and I. Procaccia, 2005, Europhys.
Lett., {\bf 72}, 943.

\bibitem[Lo et al.(2006)Lo et al.]{06LLP}
T.S. Lo, Victor S. L'vov and I. Procaccia,  2006, Phys. Rev. E.,{\bf 73}, 036308.

\bibitem[L'vov et al.(2004)L'vov et al.]{04LPPT}
V. S. L'vov, A. Pomyalov, I. Procaccia and V. Tiberkevich, 2004,
Phys. Rev. Lett.,  {\bf 92}  244503.

\bibitem[L'vov et al.(2005a)L'vov et al.]{05LPPT}
V S. L'vov, A. Pomyalov, I. Procaccia and V. Tiberkevich, 2005, Phys. Rev. E {\bf
71},  016305.

\bibitem[L'vov et al.(2005b)L'vov et al.]{Itamar1}
V. S. L'vov, A. Pomyalov, I. Procaccia and V. Tiberkevich, 2005,
Phys. Rev. Lett., {\bf 94} 174502.

\bibitem[L'vov et al.(2005c)L'vov et al.]{05LPT}
 V. S. L'vov, A. Pomyalov and V. Tiberkevich,  2005, Environmental
Fluid Mechanics, {\bf 5}, 373.

\bibitem[Lu et al.(2005)Lu et al.]{Tryggvason}
J. Lu, A. Fernandez, and G. Tryggvason, 2005, Phys. Fluids {\bf 17}, 095102.


\bibitem[Manhart(2003)Manhart]{03Man}
M. Manhart, 2003, J. NON-Newtonian Fluid Mech. {\bf 112}, 269.


\bibitem[McComb(1990)McComb]{90McComb}
W. McComb, 1990, \textit{The fluid of Physics Turbulence}, Clarendon.

\bibitem[Monin and Yaglom(1979)Monin and Yaglom]{79MY}
A. S. Monin and A. M. Yaglom, 1979, \textit{Statistical Fluid Mechanics} (MIT, 1979,
vol. 1 chapter 3).


\bibitem[Pope(2000)Pope]{Pope}
S.~B. Pope,  2000, {\it Turbulent Flows}, Cambridge University Press.

\bibitem[Ptasinski et al.(2001)Ptasinski et al.]{Nieu}
P.K. Ptasinski, F.T.M. Niewstadt, B.H.A.A. van den Brule and M.A.
Hulsen, 2001, Flow Turbul. Combust, {\bf 66}, 159.

\bibitem[Rollin(1972)Rollin]{72RS}
A.Rollin and F.A. Seyer, 1972,   Can. J. Chem. Eng, {\bf 50}, 714.

\bibitem[Rudd(1969)Rudd]{69Rud}
M.J. Rudd, 1969, Nature, {\bf 224}, 587.

\bibitem[Sanghai and Didwania(1993)Sanghai and Didwania]{stress3}
A. S. Sangani and A. K. Didwania, 1993, J. Fluid Mech. , {\bf 248} 27.

\bibitem[Sibila and Baron(2002)Sibila and Baron]{02SB}
S. Sibila and A. Baron, 2002, Phys. Fluids {\bf 14}, 1123.

\bibitem[Sreenivasan and White(2000)Sreenivasan and White]{00Sreeni}
K.R. Sreenivasan and C.M. White, 2000, J. Fluid Mech., {\bf 409} 149.

\bibitem[Toms (1949)Toms]{Toms}
B.A. Toms, 1949
\textit{in Proc. Intl. Rheological Congress Holland, 1948, p. 135}.

\bibitem[den Toonder et al.(1995)den Toonder et al.]{Dentoonder}
J. M. J. den Toonder, F. T. M. Nieuwstadt and G. D. C. KuiKen , 1995
Appl Sci Res,  {\bf 54}, 95.


\bibitem[Virk(1975)Virk]{75Virk}
P.S. Virk, 1975, AIChE J. {\bf 21}, 625.

\bibitem[Virk et al.(1996)Virk et al.]{ASME}
P. S. Virk, D. L. Wagger and E. Koury, 1996, ASME FED-
{\bf 237}, 261.


\bibitem[Virk et al.(1997)Virk et al.]{97VSW}
P.S. Virk, D.C. Sherma and D.L. Wagger, 1997, AIChE J., {\bf 43}, 3257.

\bibitem[Wagner et al.(2003)Wagner et al.]{Bonn}
C. Wagner, Y. Amarouch\`ene, P. Doyle and D. Bonn, 2003,  Europhys. Lett. {\bf 64}, 823.


\bibitem[Warholic et al.(1999)Warholic et al.]{99War}
M.D. Warholic, H. Massah and T.J. Hanratty, 1999, Exp. Fluids {\bf 27}, 461

\bibitem[Yu et al.(2001)Yu et al.]{01YKTM}
B. Yu, Y. Kawaguchi, S. Takagi and Y. Matsumoto, 2001, \textit{7th Symposium
on Smart Control of Turbulence} University of Tokyo.

\bibitem[Zagarola and Smits(1997)Zagarolla and Smits]{97ZS}
M.V. Zagarola and A.J. Smits,  1197, Phys. Rev. Lett. {\bf 78}, 239.

\bibitem[Zhang and Prosperetti(1994)Zhang and Prosperetti]{stress1}
D. Z. Zhang and A. Prosperetti, 1994, Phys. Fluids, {\bf 6}, 2956.




\end{thebibliography}
\end{document}